\documentclass[intlimits,twoside,a4paper]{article}

\usepackage{graphicx}
\usepackage[cp1251]{inputenc}

\usepackage[ukrainian,english]{babel}

\usepackage{color}
\usepackage{wrapfig}
\usepackage{caption}

\usepackage[eqsecnum]{cmpj3}

\issue{2017}{20}{2}{23501}
\doinumber{10.5488/CMP.20.23501}

\title[Phase transition in a cell fluid model]%
{Phase transition in a cell fluid model}%

\author[M.P. Kozlovskii, O.A. Dobush]{M.P. Kozlovskii, O.A. Dobush}
\address{Institute for Condensed Matter Physics of the National Academy of Sciences of Ukraine, \\ 1 Svientsitskii St., 79011 Lviv, Ukraine}

\newcommand{\no}{\nonumber}
\newcommand{\non}{\nonumber \\}

\newcommand{\be}{\begin{equation}}
\newcommand{\ee}{\end{equation}}
\newcommand{\bea}{\begin{eqnarray}}
\newcommand{\eea}{\end{eqnarray}}
\newcommand{\sli}{\sum\limits}
\newcommand{\ili}{\int\limits}
\newcommand{\lp}{\left (}
\newcommand{\rp}{\right )}

\newcommand{\vk}{\vec{k}}
\newcommand{\vl}{\vec{l}}

\newcommand{\cB}{{\cal{B}}}

\newcommand{\rhok}{{\rho_{\vec{k}}}}
\newcommand{\rhomk}{{\rho_{-\vec{k}}}}

\date{Received September 17, 2016, in final form February 5, 2017}

\begin{document}

\maketitle

\begin{abstract}
We propose a method of describing a phase transition in a cell fluid model with pair interaction potential that includes repulsive and attractive parts.
An exact representation of the grand partition function of this model is obtained in the collective variables set.
The behavior of the system at temperatures below and above the critical one is explored in the approximation of a mean-field type.
An explicit analytic form of the equation of state which is applicable in a wide range of temperatures is derived, taking into account an equation between chemical potential and density. The coexistence curve, the surface of the equation of state and the phase diagram of the cell Morse fluid are plotted.
\pacs 51.30.+i, 64.60.fd
\end{abstract}

\section{Introduction}\label{introduction}

Description of phase transitions in fluids remains a ponderable area of investigations in statistical physics both in macroscopic and microscopic scales. During many years of research in the field due to hard work of scientists it became known that the phase transition is possible either at the infinite volume or at the thermodynamic limit and is characterized by a density jump on the isotherm below the critical temperature~\cite{hansen}. Plenty of successful studies were dedicated to the explanation of the nature of the first-order phase transitions at a macroscopic level, but a resemblant theoretical description at the microscopic stage is still a question to be answered.
Most approaches to a description of phase transitions and critical phenomena in fluids are based on the following: complete scaling
ideas~\cite{PhysRevE.85.031131,Anisimov_CMP_13}; methods of
the theory of integral equations, in particular, self-consistent Ornstein-Zernike approximation
(SCOZA)~\cite{Pini_Stell_Wilding_98,Lee_Stell_Hoye_04}; perturbation series expansion such as hierarchical
reference theory~\cite{Par_Rea_12}; non-perturbative renormalization group approach~\cite{Caillol_06}; method based on the study of the behavior of the virial equation of state with extrapolated coefficients~\cite{virial};
 numerical methods and computer simulations.
There are only a few recent works devoted to finding possible ways of solving this problem in the framework of grand canonical ensemble. Among them, there is a collective variables method~\cite{yukhn_2014} used by Yukhnovskii to completely integrate the grand partition function of a system of interacting particles in the phase-space of collective variables and, therefore, investigate its behavior in the vicinity of a critical
point, or the method developed by Tang~\cite{tang} where he combined the grand canonical ensemble with density functional to explain the occurrence of a first-order phase transition in homogeneous fluids.

The objective of our investigation is microscopic description of a first-order phase transition in a cell fluid model. We do this exclusively on
 the basis of grand canonical ensemble. The model that we used as an approximation of a continuous system is quite similar to the cell gas model~\cite{rebenko_2013}.
Thermodynamic functions, correlation functions and free energy of the latter appeared to be close to the corresponding values of continuous systems if the cells size tends to zero~\cite{rebenko_2011}.

The values of parameters of the interaction potential are required to obtain numerical results. For this purpose, we chose the Morse potential as a potential of interaction. The Morse fluid has already been well studied, e.g., within the integral equation
approach~\cite{apf_11}, by molecular dynamics simulations of simple fluids~\cite{martinez}, by Monte Carlo simulations using both $NpT$ plus test particle method~\cite{okumura_00}
and the grand-canonical transition matrix method~\cite{singh}. The application of such a potential appeared to be sufficient to describe a liquid-vapor coexistence in liquid metals~\cite{singh}.

This paper is laid as follows: in section~\ref{cell_model}, the Jacobian of transition from individual to collective variables is calculated. A rigorous representation of the GPF of the cell fluid in the form of multiple integral over collective variables is obtained in section~\ref{sec3}. In section~\ref{sec4}, this expression is restricted to $\rho^4$-model and is calculated in the mean-field approximation. Sections~\ref{sec4} and \ref{sec5} are devoted to the exploration of the behavior of a system in a wide temperature range except the vicinity of the critical point, and an equation of state of the cell fluid applicable in this range is derived. Conclusions are presented in section~\ref{sec6}.

\section{A cell model for describing the behavior of a simple fluid \label{cell_model}}

 To follow our goal, we calculate the grand partition function (GPF) of the cell fluid model.

 Similar to the case of a cell gas model, the idea of the cell fluid consists in a fixed partition of the volume $V$ of a system with $N$ particles on $N_v$ congruent cubic cells, each being of a volume $v = V/N_v$~\cite{KK_arxiv}.
 The GPF of this system can be written as follows:
\be \label{GPF_1}
\Xi =\sli_{N=0}^{\infty}\frac{z^N}{N!}\int \limits_{V}(\rd x)^N\exp \left[-\frac{\beta}{2}\sli_{\vec{l}_1,\vec{l}_2\in\Lambda}\tilde U_{l_{12}} \rho_{\vl_1}(\eta) \rho_{\vl_2}(\eta)\right].
\ee
Here, $z = \re^{\,\beta \mu}$ is the activity, $\beta$ is the inverse temperature, $\mu$ is the chemical potential. In this expression, integration is made over the coordinates of all the particles in the system $\int \nolimits_{V} (\rd x)^N = \int \nolimits_{V} \rd x_1 \ldots \int \nolimits_{V} \rd x_N$, $x_i = (x_{i}^{(1)},x_{i}^{(2)}, x_{i}^{(3)}) $, $\eta = \{ x_1 , \ldots , x_N \}$ is the set of coordinates.

$\tilde U_{l_{12}}$ is the potential of interaction, the difference between two cell vectors is noted as $l_{12}= |\vec{l}_{1}- \vec{l}_{2}|$.
Each $\vec{l}_i$ takes values from a set $\Lambda$, defined as follows:
\begin{equation*} \label{Lambda}
\Lambda =\Big\{ \vl = (l_1, l_2, l_3)|l_i = c m_i;\quad m_i=1,2,\ldots,N_{1};\quad i=1,2,3;\quad N_v=N_1^3 \Big\}.
\end{equation*}
Here, $c$ is the side of a cell, $N_1$ is the number of cells along each axis, $\rho_{\,\vl}\,(\eta)$ is the occupation number of a cell
\begin{equation} \label{loc_dens}
\rho_{\,\vl}\,(\eta) = \sli_{x \in \eta} I_{\Delta_{\,\vl}(x)}\,.
\end{equation}
The characteristic functions (indicators) $I_{\Delta_{\,\vl}(x)}$
\begin{equation} \label{char_func}
I_{\Delta_{\vec{l}}(x)}=\begin{cases}
1, \qquad \text{if}\qquad x \in \Delta_{\,\vl} \, , \\
0, \qquad \text{if}\qquad x \not\in \Delta_{\,\vl}
\end{cases}
\end{equation}
identify the particles in each cubic cell $\Delta_{\,\vl} = (-c/2,c/2]^3 \subset \mathbb{R}^3$
and their contribution to the interaction of a model. This is the way to plunk all the particles and sort them to different cells.
The following expression is obvious
\begin{equation*}\label{ro_N}
    \sli_{\vl \in \Lambda} \rho_{\vec{l}}\,(\eta) = N.
\end{equation*}
Particles with different coordinates can get into the same cell, where they interact regardless of the distance between them by equally repelling each other. Interaction between the constituents of one cell and the particles hosted in another cell is expressed as the function $\tilde U_{l_{12}}$ dependent on the distance between cells. This function consists of repulsive and attractive parts, respectively
$ \tilde U_{l_{12}} = \Psi_{l_{12}} -  U_{l_{12}}$. Both $\Psi_{l_{12}}$ and $ U_{l_{12}}$ are positive.
As an example, we choose $\tilde U_{l_{12}}$ in the form of the Morse potential
\begin{equation} \label{Morse_pot}
\Psi_{l_{12}} = D \re^{-2(l_{12}- 1)/\alpha_R}, \qquad U_{l_{12}} = 2 D \re^{-(l_{12}- 1)/\alpha_R},
\end{equation}
here, $\alpha_R = \alpha / R_0$ where $\alpha$ is the effective interaction radius. The parameter $R_0$ corresponds to the minimum of the function $\tilde U_{l_{12}}$ [$\tilde U(l_{12}=1)=-D$ determines the depth of the potential well].
In terms of convenience, the $R_0$-units will be used for length measuring.

Rewriting the GPF~\eqref{GPF_1} in terms of Fourier representation gives us a sum of diagonal terms in the exponent with interaction potential
 \be \label{GPF_1a}
 \Xi = \sli_{N=0}^{\infty} \frac{z^N}{N!} \int \limits_{V} (\rd x)^N
\exp \left[-\frac{\beta}{2} \sli_{\vec{k}\in\cB_c} \tilde U(k) \hat \rho_{\vk} \hat \rho_{-\vk} \right], \qquad k = |\vec{k}|,
\ee
which is more prospective for calculation than the expression~\eqref{GPF_1}.
The variable $\hat \rho_{\vk}$ is the representation of the occupation number $\rho_{\,\vl} \,(\eta)$ in reciprocal space
\[
\hat \rho_{\vk} = \frac{1}{\sqrt{N_v}} \sli_{\vec{l}\in\Lambda}\rho_{\,\vl}\,(\eta) \re^{\ri\vec{k}\vl}.
\]
Vector $\vec{k}$ takes values from the set $\cB_c$ corresponding to one cell
\begin{equation*}\label{Bk}
    \cB_c = \Big\{  \vk = (k_{1},k_{2},k_{3})  \Big|  k_{i} = -\frac{\piup}{c}+\frac{2\piup}{c}\frac{n_{i}}{N_{1}}\,, \quad n_{i} = 1,2,\ldots,N_{1};\quad i=1,2,3;\quad  N_{v} = N_{1}^{3}   \Big\}.
\end{equation*}
The Fourier transform of the Morse potential~\eqref{Morse_pot} $\tilde U(k) = -U(k) + \Psi(k)$ is as follows:
\be \label{fourier_Morse_pot}
U(k) = U(0) \lp 1 +  \alpha_R^2 k^2 \rp^{-2}, \qquad \Psi(k) = \Psi(0) \lp 1 + \frac{\alpha_R^2 k^2 }{4} \rp^{-2},
\ee
\be \label{fourier_Morse_pot_0}
U(0) = 16 D \piup\frac{\alpha_R^3}{v} \re^{R_0/\alpha},\qquad \Psi(0) = D \piup \frac{\alpha_R^3}{v} \re^{2R_0/\alpha}.
\ee

At this stage, the GPF~\eqref{GPF_1} of $N$ interacting particles will be written in the form of $N$ integrals over their coordinates and $N_v$ integrals over the collective variables (CV) $\rho_{\vec{k}}$~\cite{yukhn_1980}
\begin{equation}\label{GPF_2}
\Xi = \sli_{N=0}^{\infty} \frac{z^N}{N!} \int \limits_{V} (\rd x)^N
\exp \left[-\frac{\beta}{2} \sli_{\vec{k}\in\cB_c} \tilde U(k) \hat \rho_{\vk} \hat \rho_{-\vk} \right]  \int (\rd \rho)^{N_v} \int (\rd \nu)^{N_v} \exp \left[ 2 \piup \ri \sum \limits_{\vec{k} \in\cB_c} \nu_{\vk} (\rho_{\vk} - \hat \rho_{\vk} )\right] .
\end{equation}
Herewith
\[ \label{div_ro}
(\rd\rho)^{N_v} = \prod \limits_{\vk\in\cB_c} \rd \rho_{\vk}\,, \qquad (\rd\nu)^{N_v} = \prod \limits_{\vk\in\cB_c} \rd \nu_{\vk}\,.
\]

Let us make an identical transformation of the potential of interaction $\tilde U(k)$ as follows:
\[ \label{pot_transf}
 \tilde U (k) = -  W(k) + \frac{\beta_{\text c}}{\beta} \chi \Psi(0).
\]
The effective interaction potential $W(k)$ is expressed as follows:
\be \label{WK2}
W(k) = U(k) - \Psi(k) + \frac{\beta_{\text c}}{\beta}  \chi \Psi(0).
\ee
Hence, $\chi$ is the real positive parameter $(\chi>0)$,
$\beta_{\text c}=1/k_{\text B} T_{\text c}$, $k_{\text B}$ is the Boltzmann constant, $T_{\text c}$ is some fixed temperature which will be defined later.
A part of the repulsive interaction $\beta_{\text c} \chi \Psi(0) >0$ is now transferred to the Jacobian of transition from individual coordinates to collective variables. This allows us to write an accurate representation of the Jacobian of transition. As we will see subsequently, it will form a new measure of distribution of density fluctuations.

As a result of the above-mentioned transformations, the GPF takes the form
\begin{align} \label{GPF_2a}
\Xi &= \sum \limits_{N=0}^{\infty} \frac{1}{N !}
    \int \limits_V (\rd x)^N
    \int (\rd \rho)^{N_v} \int (\rd \nu)^{N_v} \exp  \left[ 2 \piup \ri  \sum \limits_{\vec{k} \in\cB_c} \nu_{\vk} \big(\rho_{\vk} - \hat \rho_{\vk}  \big) \right] \nonumber \\
  &\quad \times \exp  \left[  \beta \mu \rho_0  +  \frac{\beta}{2} \sum \limits_{\vec{k}\in\cB_c} W(k) \rho_{\vk} \rho_{-\vk} - \frac{\beta_{\text c}}{2} \chi \Psi(0) \sum \limits_{\vec{k} \in\cB_c}  \hat\rho_{\vk} \hat\rho_{-\vk}  \right],
 \end{align}
here, $\exp(\beta \mu N) = \exp ( \beta \mu \rho_0 )$ and $\rho_0 = \rho_{\vk}|_{\vk = 0}\,$.

 To integrate over the coordinates $x_1,\ldots,x_N$ it is enough to consider a part of~\eqref{GPF_2a} in terms of $\rho_{\,\vl} \,(\eta)$
 \begin{equation*}\label{Strat-Habb}
   \exp \left[ - \frac{\beta_{\text c}}{2} \chi \Psi(0) \sum \limits_{\vec{k} \in\cB_c}  \hat\rho_{\vk} \hat\rho_{-\vk} \right] = \prod \limits_{l = 1}^{N_v}\exp \left[  -\frac{\beta_{\text c}}{2} \chi \Psi(0) \rho_{\,\vl}^2\, (\eta) \right]
\end{equation*}
 and make the following transformation
\begin{equation*}
    \exp \left[  -\frac{\beta_{\text c}}{2} \chi \Psi(0) \rho_{\,\vl}^2 \,(\eta) \right] = {g}_{ \Psi} \int \limits_{-\infty}^{\infty} \rd \varphi_{\,\vl} \exp \left[ -   \frac{1}{4p} \varphi_{\,\vl}^2 + \ri \varphi_{\,\vl} \rho_{\,\vl} \,(\eta) \right],  \qquad  {g}_{\Psi} = \left( 4\piup p\right)^{-\frac{1}{2}}.
\end{equation*}
 The parameter $p$ is real and positive. It is expressed by the following formula:
\be \label{p1}
p = \beta_{\text c} \chi \Psi(0)/2.
\ee

 After integrating over the coordinates $x_1, \ldots , x_n$ (see appendix~\ref{appA}) we obtain a functional representation of the GPF of the cell fluid model in the way similar to the one we did before~\cite{KD_2016}
\begin{equation}\label{GPF_3}
 \Xi = \int (\rd\rho)^{N_v} \exp \left[ \beta \mu \rho_{0} + \frac{\beta}{2} \sum \limits_{\vec{k} \in\cB_c} W(k) \rho_{\vk} \rho_{-\vk} \right]
 \prod \limits_{l=1}^{N_v} \left[ \sli_{m=0}^\infty \frac{v^m}{m!} \re^{-pm^2}\delta(\rho_{\,\vl}-m)\right].
\end{equation}

Now, the difference between the cell gas model \cite{rebenko_2013} and the cell fluid model becomes clear. Molecules of a cell gas move in space in such a way that each cell can house no more than one particle. This is somewhat similar to the lattice-gas model, which is frequently used for describing simple fluids. It is known that in the lattice-gas, only one particle can reside on each site, otherwise the site is vacant. As regards the cell fluid model, we introduce the occupation numbers of cells $\rho_{\,\vl}\,(\eta)$ which can take values $m = 0, 1, 2, \ldots\,$. This means that each cell of a cell fluid can host an arbitrary number of particles. Due to the term $\re^{-pm^2}$, the more $m$ increases the less the $m$-th term contributes to the sum in~\eqref{GPF_3}. Therefore, the probability of finding many particles in one elementary cube is very small.

\section{Functional representation of the grand partition function of the cell fluid model} \label{sec3}

At this stage, we propose two ways to deal with the calculation of the GPF~\eqref{GPF_3}. The former is presented in our recent study~\cite{KD_2016}. The latter, which in our opinion provides more transparent results, will be considered here. Firstly, let us make the Stratonovich-Hubbard transformation
\begin{equation}\label{Strat_Habb_W}
 \exp \left[  \frac{\beta}{2} \sli_{\vec{k}\in\cB_c} W(k) \rhok\rhomk \right] = g_W  \int (\rd t)^{N_v} \exp  \left[  - \frac{1}{2\beta} \sli_{\vec{k}\in\cB_c} \frac{ t_{\vk} t_{-\vk}}{W(k)} + \sli_{\vec{k}\in\cB_c} t_{\vk}\rhok \right],
\end{equation}
\[ \label{gV}
g_W = \prod\limits_{\vec{k}\in\cB_c} \left[ 2 \piup \beta W(k)\right]^{-1/2}.
\]
The variables $t_{\vec{k}}$ are complex values $t_{\vec{k}} = t_{\vec{k}}^{\text{(c)}} - \ri t_{\vec{k}}^{\text{(s)}}$, where $t_{\vec{k}}^{\text{(c)}}$ and $t_{\vec{k}}^{\text{(s)}}$
are real and imaginary parts, respectively. For an element of a phase volume, we have
\[ \label{div_t}
(\rd t)^{N_v} = \rd t_0 \prod_{\vec{k}\in\cB_c}\hspace{-1.2mm}^{\displaystyle '}\, \rd t_{\vec{k}}^{\text{(c)}} \rd t_{\vec{k}}^{\text{(s)}}.
\]
The prime above the product means that $\vec{k}>0$.

Note that the mandatory condition for transformation~\eqref{Strat_Habb_W} is $W(k)>0$. Taking into account the form of the interaction potential~\eqref{Morse_pot}, it is easy to see that the effective potential $W(k)$ \eqref{WK2} is positive when
\be \label{possible_R0A}
\ln 2 < R_0/\alpha <4 \ln 2
\ee
and $\chi > 0$. The left-hand side of inequality~\eqref{possible_R0A} is peculiar for the Morse potential, which was used \cite{girifalko,lincoln,bringas} for a description of metals (particularly, Cu, Fe, Al, Au, Na and so on).

Taking into account transformation~\eqref{Strat_Habb_W}, the GPF~\eqref{GPF_3} can be written as follows:
\be \label{GPF_4}
\Xi = g_W \int (\rd t)^{N_v} \exp \left[-\frac{1}{2\beta} \sli_{\vec{k}\in\cB_c}\frac{t_{\vec{k}}t_{-\vec{k}}}{W(k)}\right] J(\tilde{t}_{\,\vl}).
\ee

The Jacobian of transition $J(\tilde{t}_{\vec{l}})$ factors in $l$ is as follows:
\be \label{J_1}
 J (\tilde{t}_{\,\vl}) = \prod \limits_{l=1}^{N_v} \sli_{m=0}^\infty \frac{v^m}{m!} \re^{-pm^2} \int \limits_{-\infty}^{\infty} \rd\nu_{\,\vl} \,\re^{- 2 \piup \ri  \nu_{\,\vl} N} \int \limits_{-\infty}^\infty \rd\rho_{\,\vl} \exp \left[2 \piup \ri  \left(\nu_{\,\vl}+\frac{\tilde{t}_{\,\vl}}{2 \piup} \ri \right)\rho_{\,\vl}\right],
\ee
\be \label{t_tilde}
\tilde t_{\,\vl} = t_{\,\vl} + \beta \mu,  \qquad t_{\,\vl} = \frac{1}{\sqrt{N_v}} \sli_{\vec{k} \in \cB_c}t_{\vk} \re^{-\ri\vec{k}\vl}.
\ee
Obviously, integrals over $\rho_{\,\vl}$ in~\eqref{J_1} are delta-functions, which somehow remove the integration over $\nu_{\,\vl}\,$.
Consequently, we come to the following expression of the GPF
\be \label{GPF_7}
\Xi = g_W \int (\rd t)^{N_{ v}} \exp \left[  - \frac{1}{2\beta} \sli_{\vec{k}\in\cB_c}  \frac{t_{\vec{k}} t_{-\vec{k}}}{W (k)}  \right]  \prod\limits_{l=1}^{N_v} \lp  \sli_{m =0}^\infty \frac{v^m}{m!} \re^{m \tilde t_{\,\vl}} \re^{-p m^2} \rp.
\ee
The series under the product of $l$ in~\eqref{GPF_7} can be expressed as a cumulant series
\be \label{J_3}
\tilde  J_l(\tilde t_{\,\vl}) = \exp \left[ - \sli_{n=0}^\infty \frac{a_n}{n!} \tilde t_{\,\vl}^{\,n} \right].
\ee
Coefficients $a_n$ are explicit analytical functions of the parameters $p$ and $v$ (see appendix~\ref{appB}) which is in contradiction to our previous work~\cite{KD_2016}, where such coefficients were expressed in a complicated way via the integrals containing restricted cumulant series. Therefore they could be only numerically calculated. Consequently, we obtained an approximated expression of the GPF~\cite{KD_2016}.

Evidently, $p$ and $v$ appeared to be instrumental parameters in the present approach. Moreover, there is a monosemantic dependence between them (for more detail see appendix~\ref{appC}).

It is worth using a space of $\tilde t_{\vec{k}}$ to make the following calculation of the GPF. In this case, the Jacobian of transition fails to be a function of chemical potential, which actually makes calculation much easier. So, let us transit in~\eqref{GPF_7} to the variables $\tilde t_{\vec{k}}$ using~\eqref{t_tilde}, and rewrite it in the form
\begin{align} \label{GPF_8}
\Xi &= g_W \exp\left\{-\left[a_0 + \frac{\beta \mu^2}{2W(0)}\right] N_v\right\}  \int (\rd\tilde t)^{N_v} \exp \Biggl\{ \sqrt{ N_{ v}}  \left[ \frac{\mu}{W(0)} - a_1 \right] \tilde t_0 - \frac{1}{2} \sum_{\substack{\vec{k}\in \cB_c }} D(k) \tilde t_{\vec{k}} \tilde t_{-\vec{k}}  \nonumber \\
&\quad
- \sli_{n=3}^\infty \frac{a_n}{n!} N_v^{\frac{2-n}{2}}
\sum_{\substack{\vec{k}_1,\ldots\,,\vec{k}_n \\ \vec{k}_{i}\in \cB_c}} \tilde t_{\vec{k}_1} \ldots \tilde t_{\vec{k}_n}
\delta_{\vk_1+ \ldots +\vk_n}  \Biggr\}.
\end{align}
$W(0)$ is the effective potential of interaction at $|\vec{k}| = 0$ and
\[ \label{D_k}
D(k) = a_2 + \frac{1}{W(k) \beta}\, .
\]

Note that the functional representation~\eqref{GPF_8} is exact for the cell fluid model, in comparison with the visually similar expression obtained in our previous manuscript~\cite{KD_2016}, which is approximate. To derive an equation of state and physical characteristics of the system, one should calculate the GPF. Computation of~\eqref{GPF_8} would be an exact solution of the problem, but this task is rather complicated, whereas expression~\eqref{GPF_8} contains an infinite number of terms in the exponent. Thereby, we use the $\rho^4$-model approximation, which means that the cumulant series in~\eqref{GPF_8} are restricted to $n\leqslant 4$. We made sure~\cite{ma,KR_2010,KR_2011,KR_2012} that this approximation provides a qualitatively satisfactory description of a second-order phase transition with non-classical values of critical exponents for the Ising model. So, we hope that the same approximation would work for the cell fluid model as well.

\section{Calculating an explicit form of the GPF in the approximation of a mean-field type } \label{sec4}

 The expression obtained in the previous section is similar to the partition function of a three-dimensional Ising model in an external field~\cite{K_2009}. In our case, $\mu$ plays the role of the field. To calculate the GPF~\eqref{GPF_8}, we should make a change of variables
\[ \label{t_tilde_to_rok}
\tilde t_{\vec{k}} = \rho_{\vec{k}} + n_{\text c} \sqrt{N_v} \delta_{\vec{k}}\,, \qquad n_{\text c} = -\frac{a_3}{a_4}\,,
\]
which is aimed at destroying the terms proportional to the third power of the variable $\tilde t_{\vec{k}}$. As a result, we obtain the following expression
\begin{equation} \label{GPF_9}
\Xi = g_W \re^{N_v ( E_\mu - a_0 )} \int  (\rd\rho)^{N_v} \exp \Biggl[ \sqrt{ N_v} M \rho_0 - \frac{1}{2} \sli_{\vec{k}\in\cB_c} d(k) \rhok\rhomk - \frac{a_4}{24} \frac{1}{N_v} \sum_{\substack{\vec{k}_1,\ldots\,,\vec{k}_4 \\ \vec{k}_{i}\in \cB_c}} \rho_{\vec{k}_1}\ldots\rho_{\vec{k}_4} \delta_{\vec{k}_1+\ldots+\vec{k}_4}\Biggr],
\end{equation}
where
\begin{align} \label{Emu_M_a1r}
&E_\mu = -\frac{\beta W(0)}{2} ( M + \tilde{a}_1 )^2 + M n_{\text c}  + \frac{d(0)}{2}n_{\text c}^2 - \frac{a_4}{24}n_{\text c}^4\,,\non
&M = \frac{\mu}{W(0)} - \tilde{a}_1, \qquad  \tilde{a}_1 = a_1 + d(0) n_{\text c} + \frac{a_4}{6} n_{\text c}^3 \,.
\end{align}
The coefficient $d(k)$ is of the form
\be \label{dk}
d(k) = \frac{1}{\beta W(k)} - \tilde a_2\,, \qquad \tilde a_2 = \frac{a_4}{2} n_{\text c}^2 - a_2\,.
\ee

Currently, we are interested in the behavior of the cell fluid in a wide range of temperatures. The results obtained for a magnet in an external field~\cite{KPP_p_2006,K_2007} as well as for fluids~\cite{yukhn_2013,yukhn_2014} point out that the variable $\rho_0$ with $\vec{k} = 0$ ($\vec{k}$ is the wave vector of a fluctuation mode) gives a main contribution to a free energy in the range of temperatures far from the vicinity of the critical point.

The terms that include $\rhok$ with $\vk\neq 0$ fail to play a part to the average density of the system
\be \label{aver_N}
\bar n = \frac{1}{N_v}\frac{\partial\ln\Xi}{\partial\beta\mu}\,,
\ee
since integration over these variables in~\eqref{GPF_9} is not related to the chemical potential, namely a derivative of these terms with respect to the chemical potential is equal to zero. Obviously, such terms do not make the density dependent contribution to the pressure
\be \label{eq_state_gen}
P V = k_{\text B} T \ln \Xi.
\ee
Thus, let us make such a type of mean-field approximation and consider the collective variables $\rho_k$ with $\vec{k} = 0$ both with neglecting all the $\rho_k$ with $\vec{k} \neq 0$ in expression~\eqref{GPF_9}. In this approximation, the GPF is of the form
\be \label{GPF_10}
\Xi = g'_W  \re^{N_v E_\mu} \ili_{-\infty}^\infty \rd\rho_0 \exp \left[ N_v E(\rho_0)\right].
\ee
$E(\rho_0)$ is obtained using the change of variables $\rho'_0=\sqrt{N_v}\rho_0$
\be \label{Ero}
E(\rho_0)=M \rho_0 - \frac{1}{2} d(0) \rho_0^2 - \frac{a_4}{24} \rho_0^4\,.
\ee

The temperature of transition in the mean-field approximation~\cite{kadanoff,bulavin} can be determined from the following condition
\be \label{d0}
d(0) = \frac{1}{\beta_{\text c} W(0,T_{\text c})} - \tilde{a}_2= 0.
\ee
Hence, we obtain the expression for the critical temperature
\be \label{Tc}
k_{\text B} T_{\text c} = \tilde a_2 W(0,T_{\text c}),
\ee
where $W(0,T_{\text c}) = W(0)|_{T = T_{\text c}}$ is the value of the effective interaction potential $W(0)$ at the critical temperature. Since the coefficients $a_n$ of the Jacobian of transition are the functions of $p$, this parameter influences on the value of $T_{\text c}$. This dependence is important for applying the approach proposed in the present paper to particular physical systems.

Taking into account the form of the effective potential~\eqref{WK2}, it is easy to make sure that temperature dependence of $d(0)$ can be expressed as follows:
\begin{align}\label{d0gam_gam_A0}
 &d(0)  =   \tilde{a}_2 \gamma \tau, \qquad \tau = \frac{(T-T_{\text c})}{T_{\text c} }\,, \qquad
 &\gamma  =   \frac{A_0}{ A_0 + (\tau + 1)\chi } = - \frac{\tilde{U}(0)}{W(0)}\,, \qquad
 A_0 = 16 \re^{-\frac{R_0}{\alpha}} - 1.
 \end{align}
Recall that
$\tilde{U}(0) = U(0) - \Psi(0)$ is the Fourier transform of the interaction potential between the constituents of the system.

Since the value
$N_v$ in the exponent of~\eqref{GPF_10} tends to infinity in the thermodynamic limit, we can use the Laplace method to make integration in this expression (see \cite{KD_2016}) and obtain the GPF as follows:
\be \label{GPF_10a}
\Xi \simeq g'_W \exp \left[ N_v E_\mu + N_v E(\bar{\rho}_0)\right],
\ee
where the value $\rho_0 = \bar\rho_0$ can be found from the equation
\be \label{ro_eq_TmTc}
\bar\rho_{0i}^{\,3} + p_Q \bar\rho_{0i} + q_Q =0, \qquad  p_Q =  \frac{6 d(0)}{a_4}\,, \qquad q_Q = -\frac{6M}{a_4}\,.
\ee
$\bar\rho_0$ should give a maximum of $E(\rho_0)$.
Having an explicit expression of the GPF~\eqref{GPF_10a}, one can write the average density of the system using~\eqref{aver_N}
\be \label{n_bar_ae}
\bar n = \frac{\langle N \rangle}{N_v} = \frac{\partial E_\mu}{\partial\beta\mu}+ \frac{\partial E_0(\bar\rho_0)}{\partial\beta\mu} \qquad \Rightarrow \qquad  \bar n = n_g - M  + \tilde a_2 \kappa \bar\rho_0\,,
\ee
where
\be\label{ng}
n_g = - a_1 - a_2 n_{\text c} + \frac{a_4}{3} n_{\text c}^3\,,\qquad
\beta W(0) =  \frac{1}{ \tilde a_2 \kappa}\,, \qquad {\kappa} = \tau \gamma + 1,
\ee
since $\bar\rho_0=\bar\rho_0(\tau,M)$ equation~\eqref{n_bar_ae} describes a link between the density $\bar n$ and the chemical potential $M$. This equation is non-linear. It has a different number of real solutions depending on whether $\tau$ is positive or negative. At $\tau<0$, there exist three real solutions $\bar\rho_{0i}$, while for $T\geqslant T_{\text c}$, there is only one solution.

For all $\tau$, there are three possible values of $M$, which correspond to the same one $\bar n$. It is important that we can find the chemical potential as a function of the average density, but it appeared to be not an easy thing to express $M=f(\bar n,\tau)$ using~\eqref{n_bar_ae}. However, we can act in a different way to link $\bar n$ with $M$ by writing an equation that meets the condition of maximum of $E(\rho_0)$~\eqref{Ero}
\be \label{eq_M_TbTc}
M = d(0) \bar\rho_{0\text b} + \frac{a_4}{6} \bar\rho_{0\text b}^{\,3}\,,
\ee
where according to~\eqref{n_bar_ae}
\be \label{ro_TbTc_2}
\bar\rho_{0\text b} =   \frac{(\bar n - n_g + M) }{\tilde a_2 \kappa }\,.
\ee
Using the denotation
\be \label{m_TbTc}
m = M + \bar n - n_g\,,
\ee
we can reduce equation~\eqref{eq_M_TbTc} to the following form
\be \label{eq_m_TbTc}
m^3 + m p_{\text b} + q_{\text b} = 0,
\ee
where
\[ \label{pbqb}
p_{\text b} = \frac{6\kappa^2\tilde a_2^3}{a_4}\,,  \qquad q_{\text b} = - (\bar n - n_g) p_{\text b} \kappa,
\]
and its discriminant
\be \label{discrim_Db}
D_{\text b} = \lp \frac{p_{\text b}}{3}\rp^3 + \lp \frac{q_{\text b}}{2}\rp^2.
\ee

\begin{figure}[!b]
\centering
\begin{minipage}{0.49\textwidth}
\begin{center}
\includegraphics[width=0.95\textwidth]{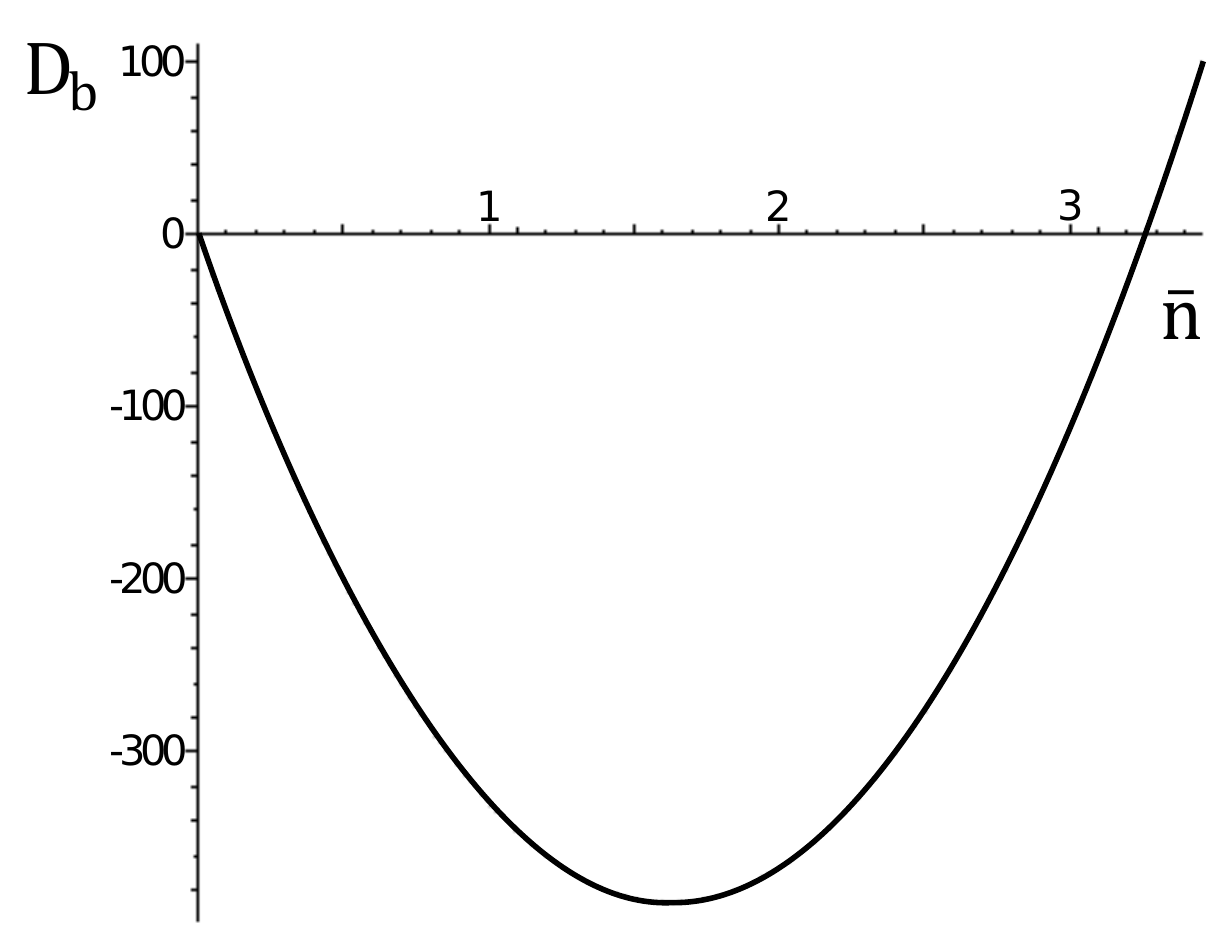}
\caption{Plot of $D_{\text b}$ as a function of average density.}\label{fig_1}
\end{center}
\end{minipage}
\begin{minipage}{0.49\textwidth}
\begin{center}
\includegraphics[width=0.95\textwidth]{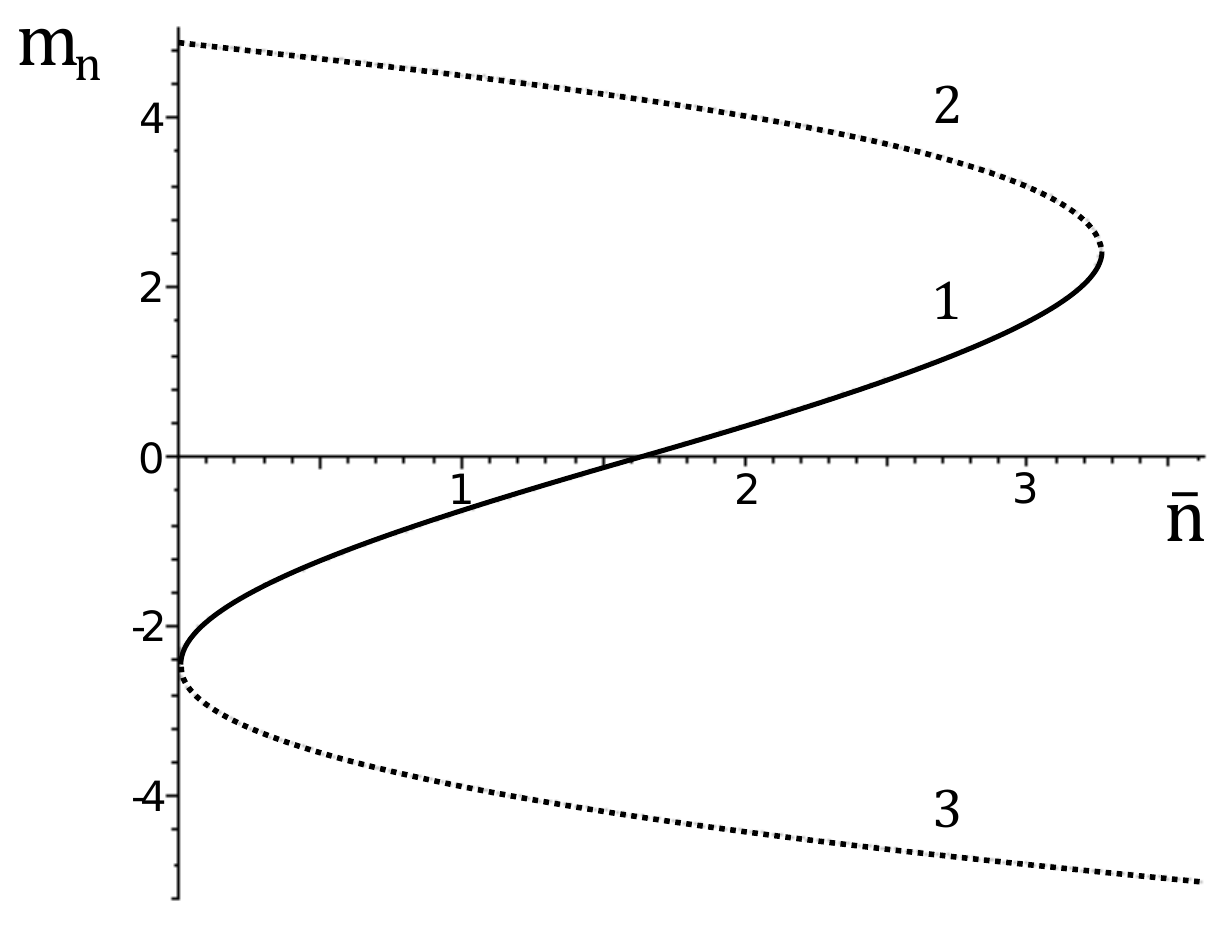}
\caption{Solutions $m_n$ as functions of average density $\bar n$: $m_1$ --- curve~1, $m_2$ --- curve~2, $m_3$ --- curve~3.}\label{fig_2}
\end{center}
\end{minipage}
\end{figure}

Since $p_{\text b}<0$, the value of $D_{\text b}$ can be either positive or negative (see figure~\ref{fig_1}). This fact shows the influence on the number of real solutions of equation~\eqref{eq_m_TbTc}. If $D_{\text b}<0$, the equation~\eqref{eq_m_TbTc} has three real solutions. If $D_{\text b} = 0$, there are two values of density $\bar{n}$. In case of $D_{\text b} > 0$, the equation has a single solution describing a decrease of $\bar{n}$ with an increase of the chemical potential, and this is an unphysical behavior.

In the ranges of density $0 \leqslant \bar{n} \leqslant \bar{n}_{\text{max}}$,
$D_{\text b}$ is negative, as it is plotted in figure~\ref{fig_1}.

Solutions of equation~\eqref{eq_m_TbTc} are of the form
\be \label{solutions_m123}
m_{1} = 2\sqrt{\frac{2\kappa^2\tilde a_2^3}{a_4}} \sin \frac{\alpha_{\text b}}{3}\,,\quad
m_{2} = - 2\sqrt{\frac{2\kappa^2\tilde a_2^3}{a_4}} \sin \lp \frac{\alpha_{\text b}}{3} + \frac{\piup}{3}\rp,\quad
m_{3} = - 2\sqrt{\frac{2\kappa^2\tilde a_2^3}{a_4}} \sin \lp \frac{\alpha_{\text b}}{3} - \frac{\piup}{3}\rp,
\ee
where
\[ \label{alpha_b}
 \alpha_{\text b} = \arcsin \left[ \sqrt{\frac{9 a_4}{8 \tilde{a}_2^3}}(\bar n - n_g)\right].
\]

Note that the solution $m_1$ in~\eqref{solutions_m123} (see curve~1 in figure~\ref{fig_2}) is equitable for some (bounded) range of values of the chemical potential $M=f(\bar n,\tau)$
 \be \label{M_n_bar}
M = m_1 - (\bar n - n_g).
\ee

\begin{figure}[!b]
\centering
\begin{minipage}{0.49\textwidth}
\begin{center}
\includegraphics[width=0.95\textwidth]{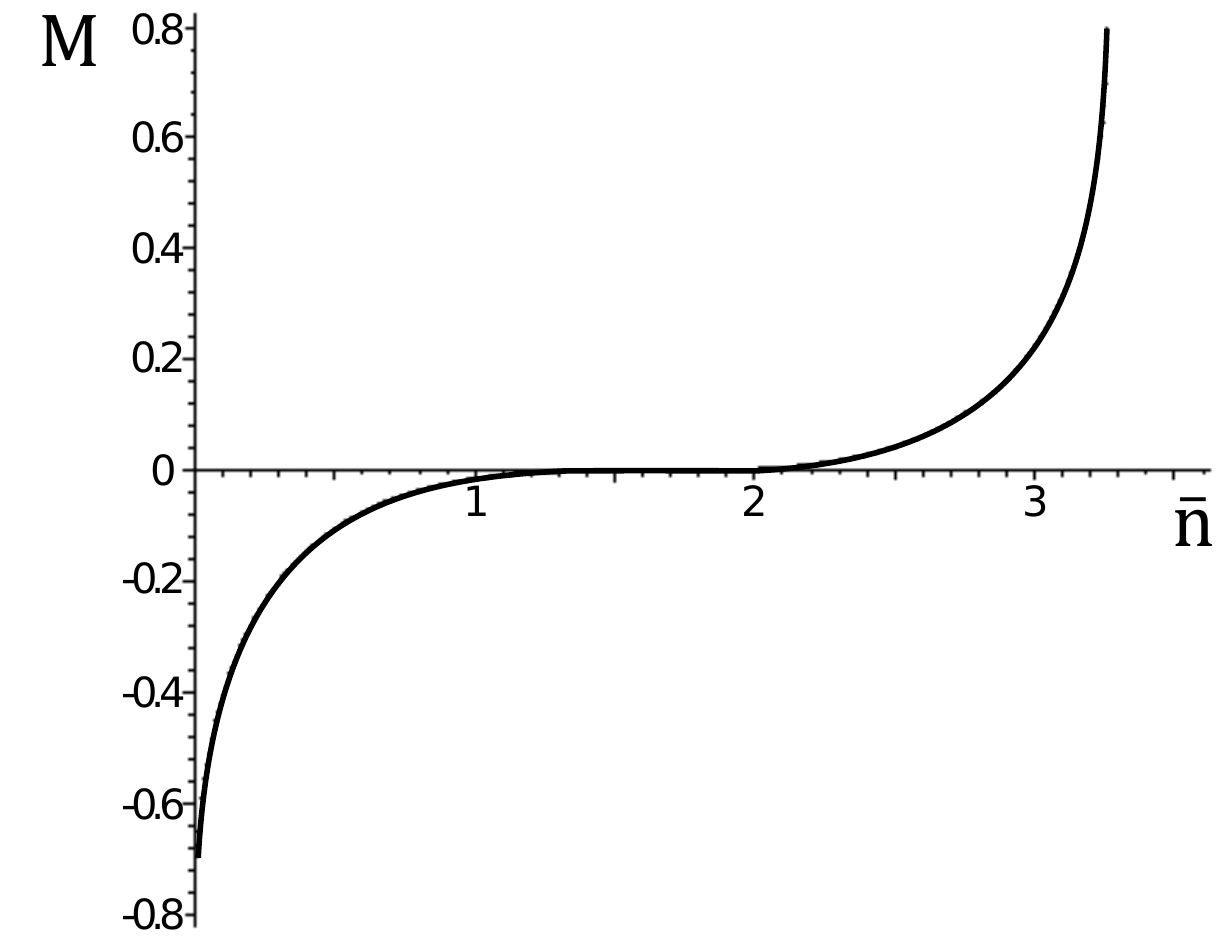}
\caption{Plot of the chemical potential as a function average density.}\label{fig_3}
\end{center}
\end{minipage}
\begin{minipage}{0.49\textwidth}
\begin{center}
\includegraphics[width=0.95\textwidth]{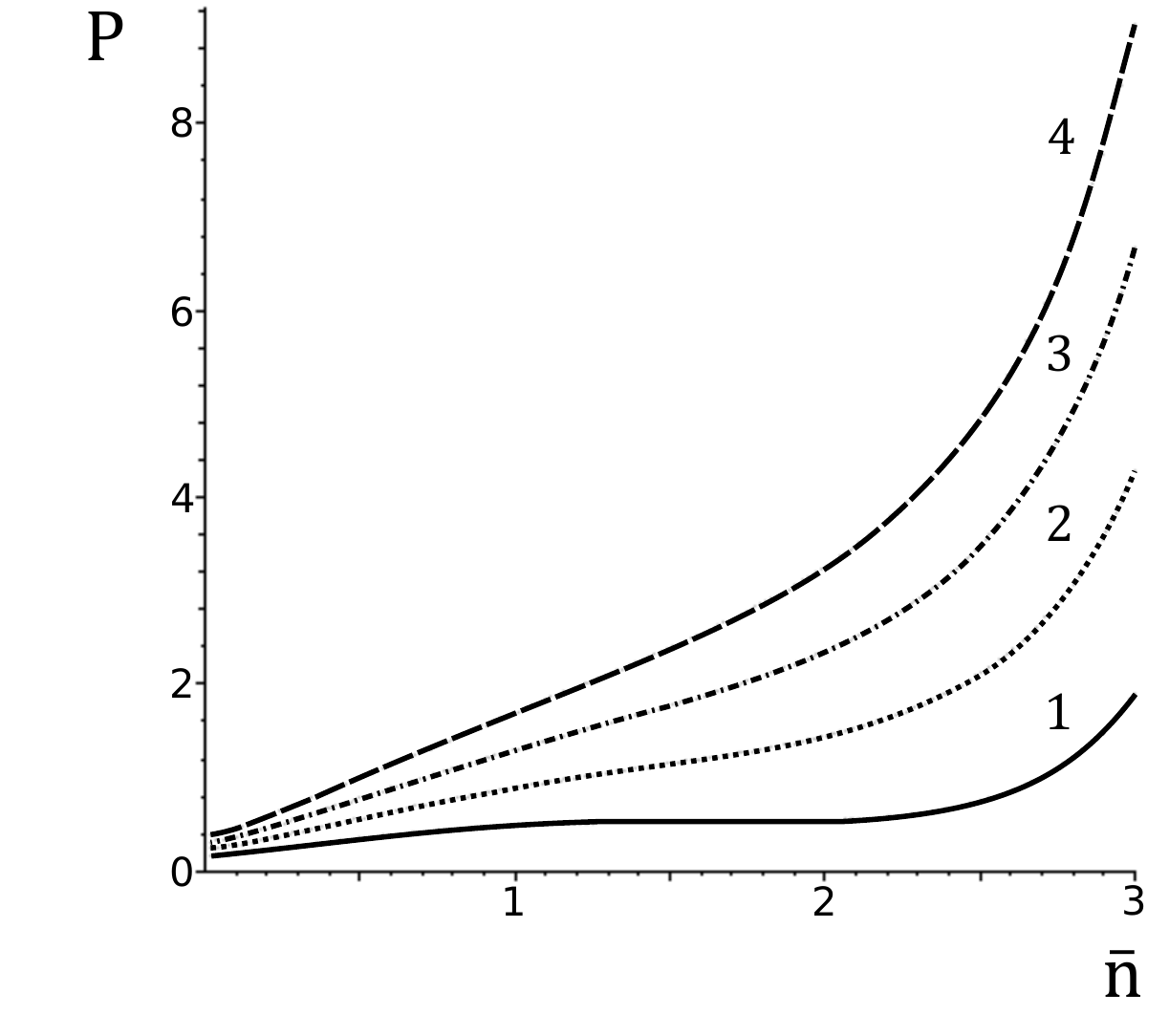}
\caption{Plot of the pressure as a function of average density at different temperatures ($\tau=0$ --- curve~1, $\tau=0.5$ --- curve~2, $\tau=1$ --- curve~3, $\tau=1.5$ --- curve~4).}\label{fig_4}
\end{center}
\end{minipage}
\end{figure}

At the parameters given in~\eqref{pAzchi_values},
the chemical potential take values
$-0.7 < M < 0.7$ (figure~\ref{fig_3}).
If $M$ falls out of this interval, there exist two other solutions ($m_2$ and $m_3$) of equation~\eqref{M_n_bar}, but they do not reflect a physical matter of the phenomenon.

Going back to the calculation of the GPF~\eqref{GPF_10a}, the case $T>T_{\text c}$ means a single real solution of~\eqref{ro_eq_TmTc}
\be\label{ro_solution_TbTc}
\bar\rho_{0\text b} = \lp \frac{3M}{a_4} + \sqrt Q \rp^{\frac{1}{3}} + \lp \frac{3M}{a_4} - \sqrt Q \rp^{\frac{1}{3}}, \qquad Q = \left[ \frac{2 d(0)}{a_4}\right]^3 + \left( - \frac{3M}{a_4}\right)^2.
\ee

 Then, the equation of state or in other words the pressure as a function of temperature and density is of the form
\begin{equation} \label{eq_state_TbTc_2}
\frac{P v}{ k_{\text B} T_{\text c}} = (\tau+1) \Bigg[  f - \frac{(M + \tilde a_1 )^2}{2 \kappa \tilde a_2 }  + \frac{d(0)}{2(\kappa\tilde a_2 )^2} \lp \bar n - n_g + M + \kappa \tilde a_2 n_{\text c} \rp ^2 + \frac{a_4}{8 (\kappa\tilde a_2)^4} \lp \bar n - n_g + M + \kappa \tilde a_2 n_{\text c} \rp^4  \Bigg],
\end{equation}
\[ \label{f}
f =  \frac{1}{N_v} \ln g'_W - a_0 + \frac{a_4}{24} n_{\text c}^4.
\]
Now, considering $T=T_{\text c}$ in~\eqref{eq_state_TbTc_2}, it is easy to write an explicit dependence of pressure on density at the critical temperature
\be \label{eq_state_Tc_main}
\frac{P v}{ k_{\text B} T_{\text c}} = f_{\text c}
 + \frac{M_0}{4 \tilde a_2} \left( M_0 + 3 \bar n\big|_{T=T_{\text c}} - n_g \right),
\ee
\[ \label{fc}
 f_{\text c} = \frac{1}{N_v}\ln g'_W - a_0 + \frac{a_4}{24} n_{\text c}^4 - \frac{(a_1 + a_4 n_{\text c}^3 / 6)^2}{2 \tilde a_2}\,.
\]
The following formula expresses the chemical potential $M_{0}$ as a function of density
\[ \label{Mc_solution}
M_{0} =  2\lp \frac{2\tilde a_2^3}{a_4}\rp^{1/2} \sin \left( \frac{ 1}{3}\alpha_{\text b}\big|_{T=T_{\text{c}}}  \right) - \left(\bar n\big|_{T=T_{\text c}}-n_g\right).
\]
Index $0$ denotes that $M_0$ corresponds to $T=T_{\text c}$.

The pressure $P = P( \bar{n} )$ expressed by~\eqref{eq_state_TbTc_2} and $P|_{T=T_{\text c}}= P|_{T=T_{\text c}}( \bar{n} )$ expressed by~\eqref{eq_state_Tc_main} are plotted in figure~\ref{fig_4}.

\section{Behavior of the system at temperatures $T<T_{\text c}$} \label{sec5}

The equation of state in case of $T<T_{\text c}$ can be written in the form
\be \label{eq_state_TmTc_1}
\frac{P v}{k_{\text B} T} = \frac{1}{N_v} \ln g'_W + E_\mu + M \bar\rho_{0i} - \frac{1}{2} d(0) \bar\rho_{0i}^{\,2} - \frac{a_4}{24} \bar\rho_{0i}^{\,4}\,,
\ee
where $E_\mu$ is defined in~\eqref{Emu_M_a1r}, and values $\bar\rho_{0i}$, $i=1,2,3$, are the three solutions of equation~\eqref{ro_eq_TmTc} \cite{KD_2016}
\be \label{solutions_ro123}
\bar\rho_{01} = 2 \rho_{0\text r}\cos \frac{\alpha_{m}}{3}\,,\qquad
\bar\rho_{02} = - 2 \rho_{0\text r}\cos \lp \frac{\alpha_{m}+\piup}{3} \rp,\qquad
\bar\rho_{03} = - 2 \rho_{0\text r}\cos \lp \frac{\alpha_{m}-\piup}{3} \rp,
\ee
where
\be \label{ror}
\rho_{0\text r} =  \sqrt{- \frac{2 d(0)}{a_4}}\,, \qquad \alpha_{m} = \arccos \frac{M}{M_q}\,.
\ee
$M_q$ is the value of the chemical potential $M$ when the discriminant $Q$ of equation~\eqref{ro_eq_TmTc} is equal to zero
\be \label{Mq}
M_q = \sqrt{-\frac{8[d(0)]^3}{9 a_4}}\,.
\ee
Note that the region of values of $M$ is restricted to the interval $|M| \leqslant M_q.$
For all $|M| \leqslant M_q$, we have $Q<0$, so there exist three real solutions of equation~\eqref{ro_eq_TmTc} in this interval of $M$.

In case of $|M| > M_q$,
we have a single root of equation~\eqref{ro_eq_TmTc}, as well as in case of $T>T_{\text c}$~\eqref{ro_solution_TbTc}.

At $M=-M_q$ we have $Q=0$, so using~\eqref{ro_solution_TbTc} we obtain
\be \label{roqm}
\bar\rho_{0q}^{\,(-)} = 2 \lp - \frac{3M_q}{a_4}\rp^{\frac{1}{3}} \equiv - 2 \rho_{0\text r}\,,
\ee
and at $M=M_q$ using~\eqref{ro_solution_TbTc} we can find
\be \label{roqp}
\bar\rho_{0q}^{\,(+)} = 2 \lp  \frac{3M_q}{a_4}\rp^{\frac{1}{3}} \equiv 2 \rho_{0\text r}.
\ee
Let us analyse an asymptotics of the solutions~\eqref{solutions_ro123} at $|M|=M_q$. Here, at $M=-M_q$, we have $\cos \alpha_q^{(-)}=-1$, that is why $\alpha_{m}^{(-)}=\piup$ and
\bea \label{ro123m}
&&
\bar\rho_{01}^{\,(-)} = 2 \rho_{0\text r}\cos \frac{\piup}{3} = \rho_{0\text r}\,,\non
&&
\bar\rho_{02}^{\,(-)}  = -2 \rho_{0\text r}\cos \lp \frac{2\piup}{3} \rp = \rho_{0\text r}\,,\\
&&
\bar\rho_{03}^{\,(-)} = - 2 \rho_{0\text r}\cos 0 = - 2 \rho_{0\text r}.\no
\eea
In case of $M=M_q$  $\cos\alpha_{m}^{(+)}=1$, or $\alpha^{(+)}_q=0$, then
\bea \label{ro123p}
&&
\bar\rho_{01}^{\,(+)} = 2 \rho_{0\text r}\cos 0 = 2 \rho_{0\text r}\,,\non
&&
\bar\rho_{02}^{\,(+)} = - 2 \rho_{0\text r}\cos \frac{\piup}{3} = - \rho_{0\text r}\,,\\
&&
\bar\rho_{03}^{\,(+)} = - 2 \rho_{0\text r}\cos \lp - \frac{\piup}{3}\rp = - \rho_{0\text r}.\no
\eea
Thereby, for $M=-M_q$ we have $\bar\rho_{0q}^{\,(-)}=-2 \rho_{0\text r}$ defined in~\eqref{roqm} coinciding with $\bar\rho_{03}^{\,(-)}$ expressed in~\eqref{ro123m}.

In case of $M=M_q$ $\bar\rho_{0q}^{\,(+)}=2\bar\rho_{03}$ from~\eqref{roqp}, which is the same as $\bar\rho_{01}^{\,(+)}$ expressed in~\eqref{ro123p}.

\begin{figure}[!t]
\vspace{-5mm}
\begin{centering}
\includegraphics[width=250pt]{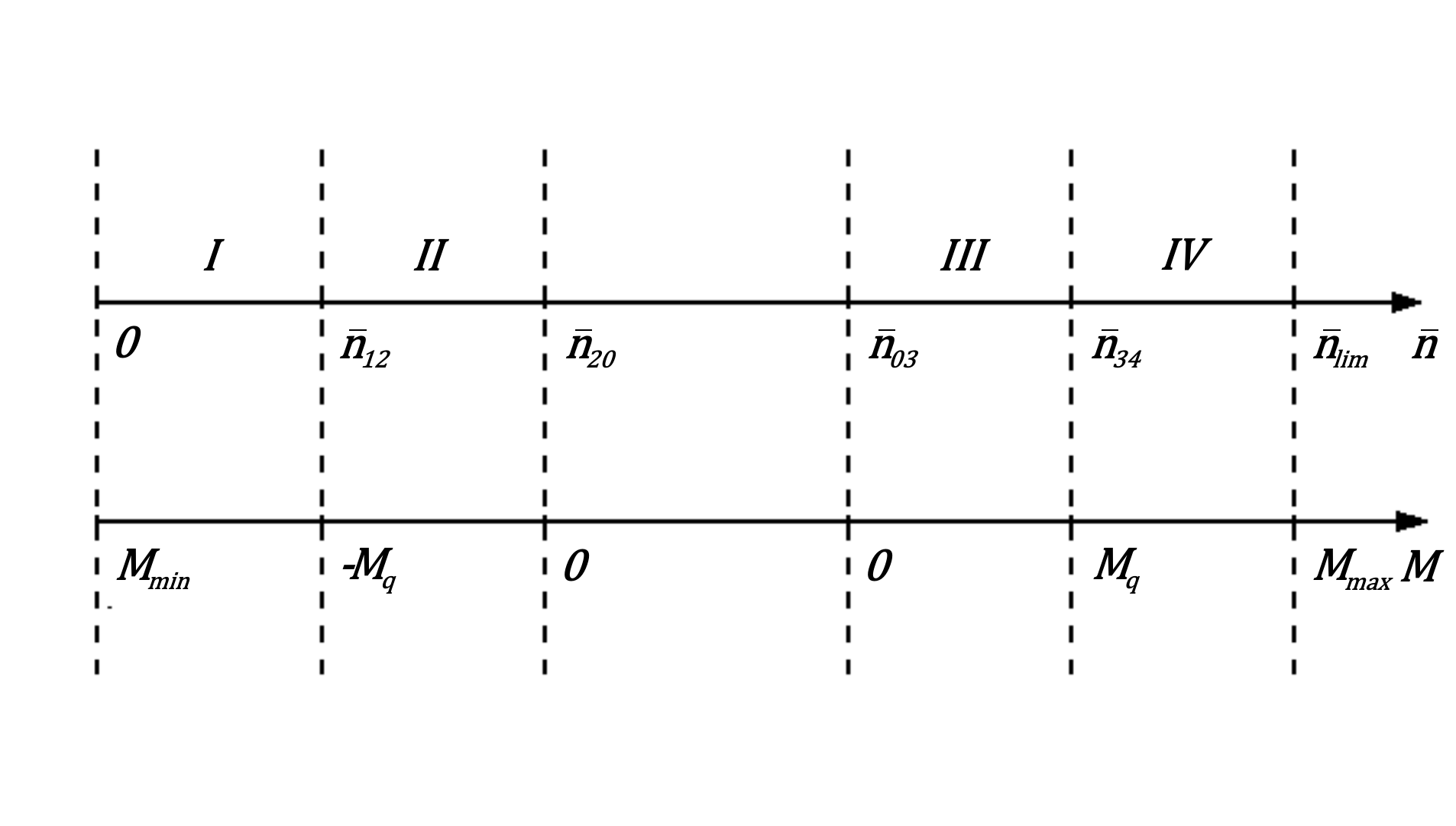}
\vspace{-5mm}
\caption{Regions of values of the chemical potential and the corresponding values of the average density at temperature below the critical one.}\label{fig_7}
\end{centering}
\end{figure}

Now we can draw the following conclusion. An increase of the chemical potential from $-\infty$ to $-M_q$ corresponds to a single solution of equation~\eqref{ro_eq_TmTc}, which is given in~\eqref{ro_solution_TbTc} (region~I in figure~\ref{fig_7}). For $-M_q<M<0$, this solution transits to $\bar\rho_{03}^{\,(-)}$ expressed in~\eqref{ro123m} and remains to be applicable until $M=-0$ (region~II) where $\cos \alpha_0^{(-)}=0$ or $\alpha_0^{(-)}=\frac{\piup}{2}$
\be \label{lim_ro3m}
\lim\limits_{M\rightarrow -0} \rho_{03}^{(-)} = \rho_0^{(-)} = - 2 \rho_{0\text r} \cos \lp \frac{\piup}{6} - \frac{\piup}{3} \rp = - \sqrt{ 3} \rho_{0\text r}.
\ee
On the other hand, a decrease of $M$ from $+\infty$ to $M_q$ corresponds to a single solution~\eqref{ro_solution_TbTc} (region~IV), which at $M=M_q$ transits to the solution $\bar\rho_{01}^{\,(+)}$ expressed in~\eqref{ro123p} and remains to be applicable until $M=+0$ (region~III):
\be \label{lim_ro1p}
\lim\limits_{M\rightarrow +0} \rho_{01}^{(+)} = \rho_{0}^{(+)} =  2 \rho_{0\text r} \cos \lp \frac{\piup}{6} \rp = \sqrt 3 \rho_{0\text r}.
\ee

\begin{figure}[!t]
\begin{centering}
\includegraphics[width=190pt]{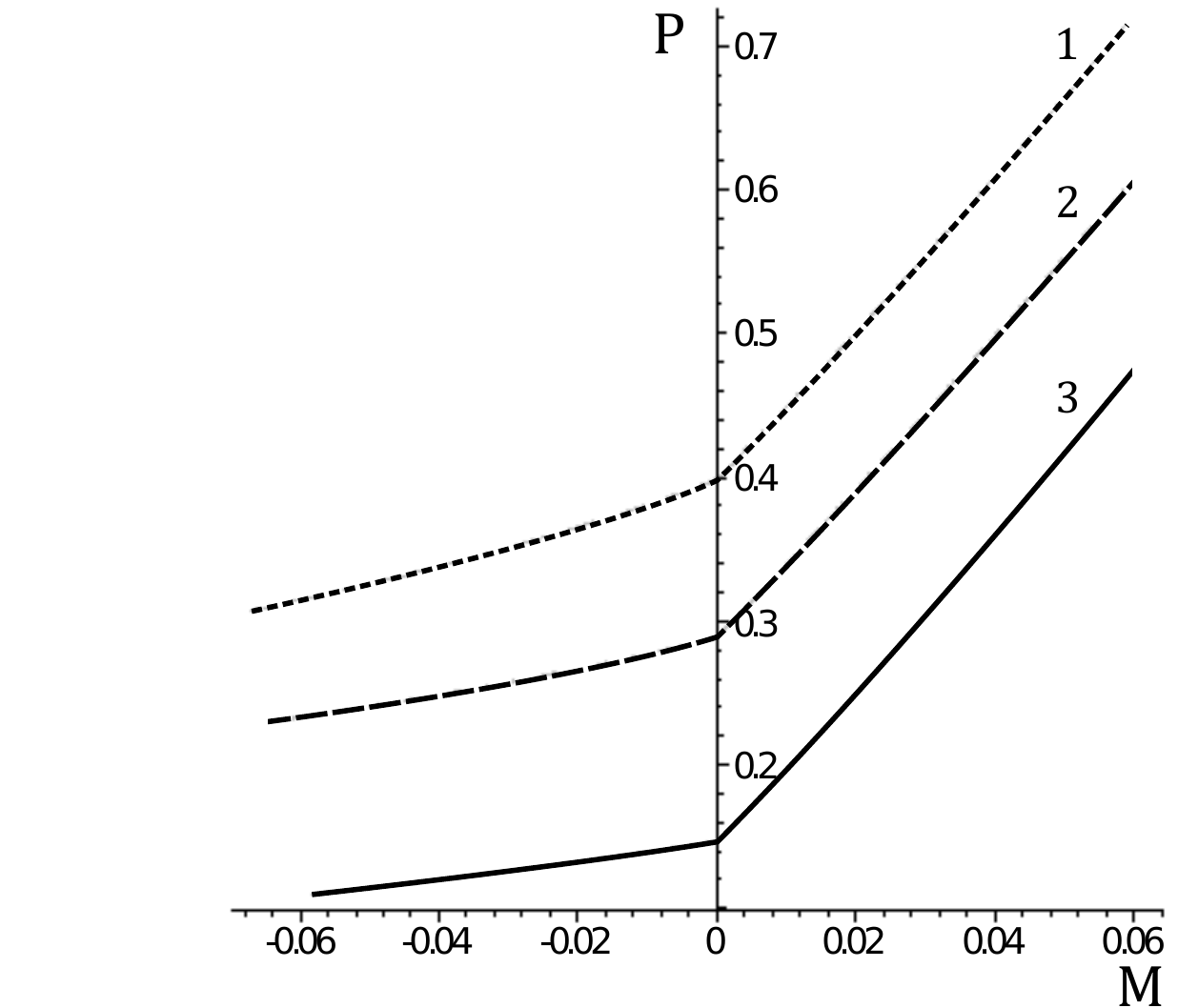}
\caption{Plot of the pressure $P$ as a function of chemical potential and
temperature at $T<T_{\text c}$ ($\tau=-0.1$ --- curve~1, $\tau=-0.2$ --- curve~2, $\tau=-0.3$ --- curve~3).}\label{fig_6}
\end{centering}
\end{figure}

Thus, the equation of state at $T<T_{\text c}$ can be written in the form of several terms using the Heaviside functions $\theta$ in accord with the values of the chemical potential $M$
\begin{align} \label{eq_state_TmTc_3}
\frac{P v}{k_{\text B}T} &= -\frac{1}{N_{ v}} \ln g'_W + E_\mu + E  (\bar\rho_0) \left[ \Theta (  -M - M_q  ) + \Theta (  M - M_q  )\right]+ E  (\bar\rho_{03}) \Theta (  -M ) \Theta ( M + M_q ) \non
&\quad  + E  (\bar\rho_{01}) \Theta( M ) \Theta ( M_q  - M ).
\end{align}
This information is sufficient to plot the function $P=P(\tau,M)$. The corresponding curves are shown in figure~\ref{fig_6}. There we can see lines broken at the point $M=0$. In case of $T \geqslant T_{\text c}$ such curves are smooth.

However, the plot of the pressure as a function of density and temperature is more informative. To get this dependence, we should express the chemical potential as a function of density according to the interrelation~\eqref{n_bar_ae} and use the obtained expression in~\eqref{eq_state_TmTc_3}. For this purpose, let us use~\eqref{M_n_bar} and compare the behavior of $\bar\rho_0$ as a function of chemical potential on the one hand and, on the other hand, as a function of density.
Consider each of the regions depicted in figure~\ref{fig_7}.

{\textbf{Region I}}, $M\leqslant - M_q$. At $M=-M_q$, we have expression~\eqref{roqm} for $\bar\rho_0$, where $\rho_{0\text r}$ is a function of temperature~\eqref{ror}. On the other hand, there is equality~\eqref{n_bar_ae}
\be
\bar n = n_g - M + \bar\rho_{0} \kappa\tilde a_2\,,
\ee
which links the average density $\bar n$ with the chemical potential (in the present case $M=-M_q$) and the value $\bar\rho_{0} = -2\rho_{0\text r}$.
Thereby, according to~\eqref{n_bar_ae} we can find the density $\bar n_{12}$ (the point of transition between region~I and region~II), which corresponds to $M=-M_q$
\be \label{n12}
\bar n_{12} = n_g - 2 \kappa \tilde a_2 \rho_{0\text r} + \frac{a_4}{3} \rho^3_{0\text r}.
\ee

{\textbf{Region II}}, $-M_q\leqslant M \leqslant -0$. At $M = -0$, we have $\bar\rho_{03}=-\sqrt 3 \rho_{0\text r}$. A boundary value of density which is pertinent to $\bar n_{20}=\lim \limits_{M\rightarrow -0} \bar n$ is of the form
\be \label{n20}
\bar n_{20} = n_g - \sqrt 3 \kappa \tilde a_2 \rho_{0\text r}.
\ee

{\textbf{Region III}}, $0\leqslant M \leqslant M_q$. This region starts from $M=0$, where $\bar\rho_{01}=\sqrt 3 \rho_{0\text r}$ and respectively
\be \label{n03}
\bar n_{03} = n_g + \sqrt 3 \kappa \tilde a_2 \rho_{0\text r}\,,
\ee
and ends up at $M=M_q$, which takes the consequences with $\bar\rho_{01}=2\rho_{0\text r}$, therefore,
\be \label{n34}
\bar n_{34} = n_g + 2 \kappa\tilde a_2 \rho_{0\text r} - \frac{a_4}{3} \rho^3_{0\text r}.
\ee

\begin{figure}[!t]
\centering
\begin{minipage}{0.49\textwidth}
\begin{center}
\includegraphics[width=0.87\textwidth]{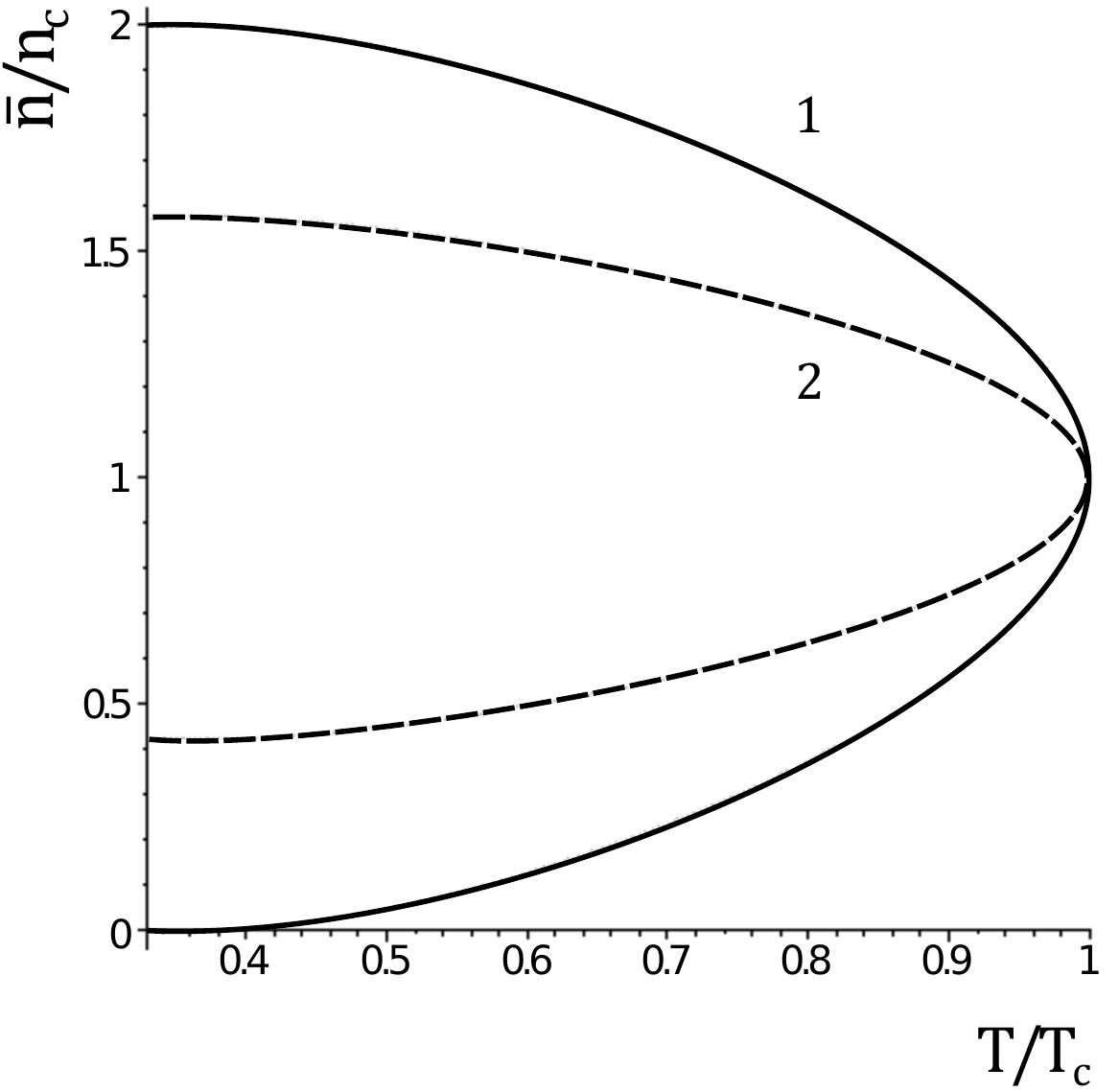}
\caption{Plot of the temperature of phase transition as a function of density  (binodal --- curve~1, spinodal --- curve~2).}\label{fig_8}
\end{center}
\end{minipage}
\begin{minipage}{0.49\textwidth}
\begin{center}
\includegraphics[width=0.95\textwidth]{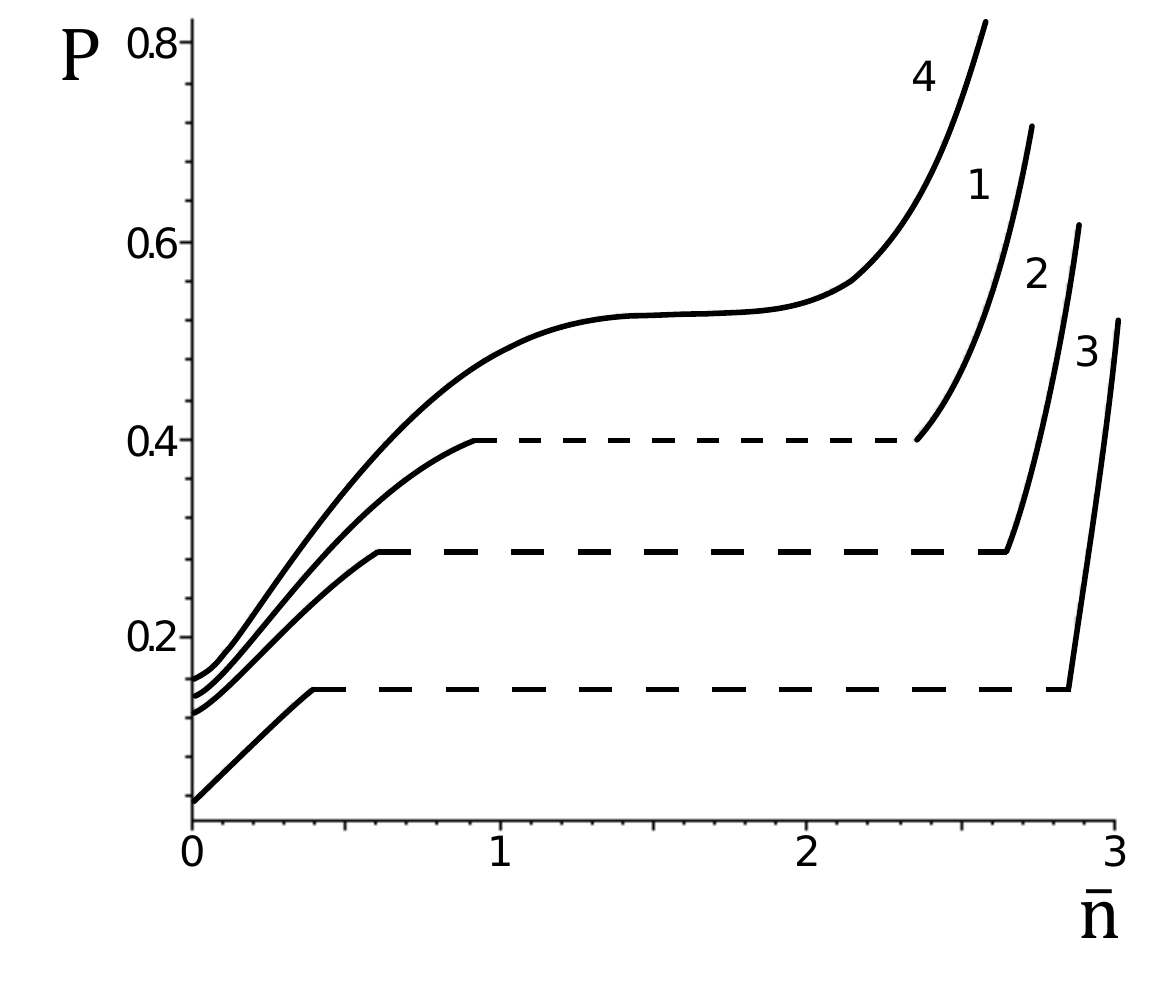}
\caption{Plot of the pressure $P$ as a function of density and temperature $\tau = 0$ --- curve~4, $\tau = -0.1$ --- curve~1, $\tau = 0.2$ --- curve~2, $\tau = -0.3$ --- curve~3.}\label{fig_9}
\end{center}
\end{minipage}
\end{figure}

{\textbf{Region IV}}, $M>M_q$. This region starts from $\bar n_{03}$~\eqref{n03}, which decreases up to some fixed value~$n_{\text{lim}}$. The value $n_{\text{lim}}$ is bounded. The plot of $\bar n_{20}$ and $\bar n_{03}$ as functions of temperature is shown in figure~\ref{fig_8} (the binodal curve). Maximal value of $M$ corresponds to $n_{\text{lim}}= 3.26$. In case of $\bar n>n_{\text{lim}}$, the solution $m_1$ in~\eqref{M_n_bar} should be replaced by $m_3$. However, this solution is non-physical, because it gives a decrease of the chemical potential with an increase of the density.

Having a temperature dependence of the boundary densities
 $\bar n_{12}$, $\bar n_{20}$, $\bar n_{03}$ and $\bar n_{34}$, one can rewrite the equation of state~\eqref{eq_state_TmTc_3} in terms of $(\tau, \bar n)$. For this purpose, it is sufficient to use the formula~\eqref{M_n_bar} in equation~\eqref{eq_state_TmTc_3}. This formula gives a dependence of the chemical potential on temperature and the average density. The boundary densities described above should be substituted for corresponding values of the chemical potential in Heaviside functions $\Theta(M_i-M_q)$ in~\eqref{eq_state_TmTc_3}. As a result, the equation of state is expressed as follows:
\begin{align} \label{eq_state_TmTc_4}
\frac{P v}{k_{\text B}T} &= - \frac{1}{N_v} \ln g'_W + E_\mu(\bar n) + E_1(\bar\rho_{03}) \Theta(\bar n_{12}-\bar n)
+ E_2(\bar\rho_{03})\Theta(\bar n - \bar n_{12}) \Theta(-\bar n + \bar n_{20}) \nonumber \\
&\quad+ E_3(\bar\rho_{01}) \Theta(\bar n - \bar n_{03}) \Theta(\bar n_{34} - \bar n) + E_4(\bar\rho_{01}) \Theta(\bar n - n_{34}).
\end{align}
$E_\mu$ is determined by the formula~\eqref{Emu_M_a1r}. Functions $E_n(\bar\rho_0)$ are of the following form
\be \label{Enro}
E_n(\bar\rho_{0n}) = M \bar\rho_{0n} - \frac{ d(0)}{2} \bar\rho_{0n}^{\,2} - \frac{a_4}{24} \bar\rho_{0n}^{\,4} \,,
\ee
where $\bar\rho_{0n}$ is either $\bar\rho_{01}$ [see~\eqref{solutions_ro123}] for $E_3(\bar\rho_{01})$ and $E_4(\bar\rho_{01})$, or $\bar\rho_{03}$ [see~\eqref{solutions_ro123}] for $E_1(\bar\rho_{03})$ and $E_2(\bar\rho_{03})$.

The pressure $P$ as a function of density and temperature is plotted in figure~\ref{fig_9}. It is easy to see that in the temperature range $\tau<0$, there is a jump of density depicting a coexistence curve.
Numerical results show that one can observe a liquid phase at $T<T_{\text c}$ in the density region $\bar n\leqslant n_{\lim}$.

\begin{figure}[!t]
\centering
\begin{minipage}{0.49\textwidth}
\begin{center}
\includegraphics[width=0.95\textwidth]{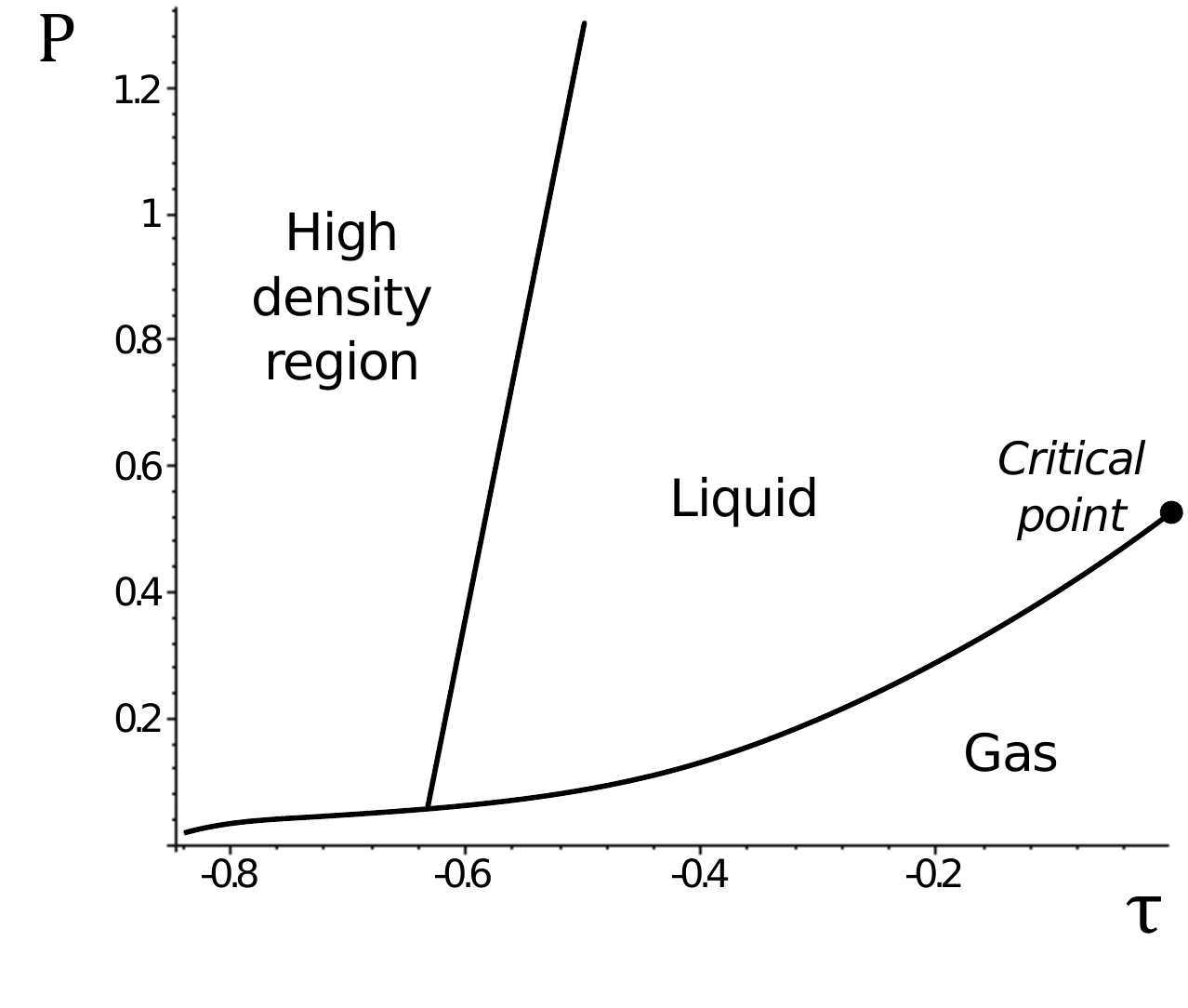}
\caption{Phase diagram.}\label{fig_10}
\end{center}
\end{minipage}
\begin{minipage}{0.49\textwidth}
\begin{center}
\includegraphics[width=0.99\textwidth]{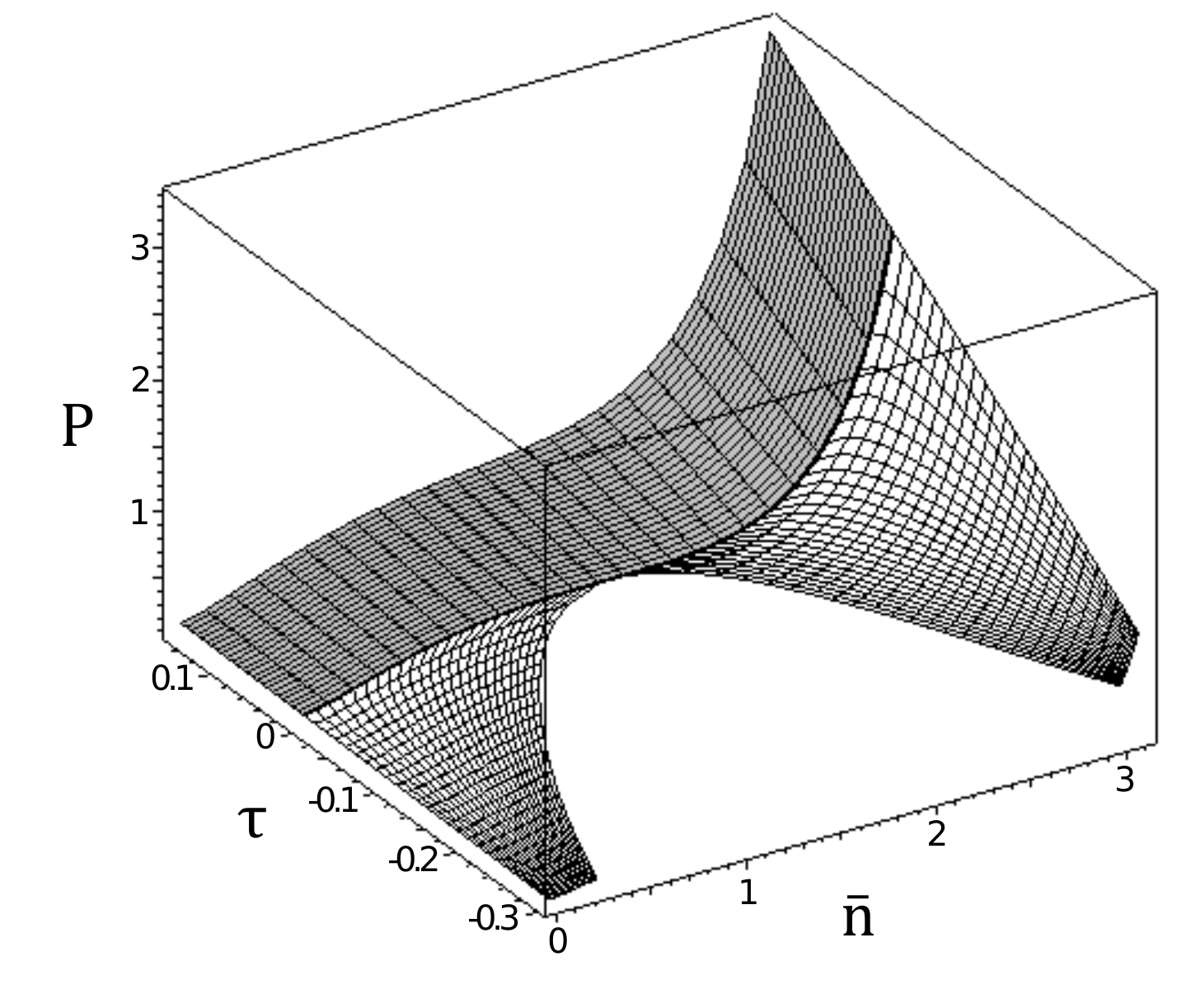}
\caption{Surface of pressure as a function of average density and temperature.}\label{fig_11}
\end{center}
\end{minipage}
\end{figure}

On the phase diagram (figure~\ref{fig_10}), one can see the line of some limited pressure $P_{\text{lim}}= \lim \limits_{\bar n \rightarrow n_{\text{lim}}} P (\bar n)$ connected to the maximal value of the fluid density $n_{\text{lim}}$. This boundary value depends on the microscopic parameters of the system. In other words, it takes a specific value for each system. The line of maximal density may appear as a consequence of using the $\rho^4$-model [see~\eqref{GPF_9} which is the approximation of~\eqref{GPF_8}]. This line sets the range of densities $0 \leqslant n \leqslant n_{\text{lim}}$ applicable to the description of a fluid behavior in this manuscript.

Black line between the gray and the white zones of the surface of pressure as a function of temperature and average density (figure~\ref{fig_11}) corresponds to the range of temperatures close to the critical $T_{\text c}$, where the mean-field approximation is not applicable.

\section{Conclusions} \label{sec6}

To describe a first-order phase transition in the cell fluid model, we deal with the calculation of the grand partition function.
A rigorous functional representation of the GPF~\eqref{GPF_8} is obtained, which is reduced to the $\rho^4$-model~\eqref{GPF_9}. This expression should be calculated in two stages. The former is described in this paper. It covers a wide range of temperatures except the critical point. The latter stage concerns a narrow vicinity of the critical point. It requires the consideration of contribution from all the variables~$\rho_{\vec{k}}$ with $\vec{k} \neq 0$ within the field-theoretical approach.

In the present work we were restricted to the use of a simplest approximation, which is called the approximation of a mean-field type. Thereby, determination of the GPF consists in the computation of a single integral using the Laplace method. As a result, we obtained an explicit form of the GPF as a function of temperature and chemical potential in this approximation. The pressure as a function of temperature and average density [equations~\eqref{aver_N} and~\eqref{eq_state_gen}] is found. Figure~\ref{fig_7} is important in this respect. It shows a correspondence between the regions of chemical potential and density at temperatures below the critical one $T_{\text c}$. The critical temperature is defined by~\eqref{Tc}. Referring to figure~\ref{fig_6} it is easy to see that transition of the effective chemical potential $M$ across the value $M=0$ changes the behavior of the pressure $P$ at $T<T_{\text c}$. Here, we have $\lim \limits_{M \rightarrow -0} P = \lim \limits_{M \rightarrow +0} P$, although the derivatives of pressure on the left-hand side and on the right-hand side from $M=0$ are different. The latter fact causes a jump of density starting from $n_{20}$ to $n_{03}$ (see figure~\ref{fig_7}). Each of these values are functions of temperature. In case of $T \geqslant T_{\text c}$, we do not observe such a jump (formally $n_{20} = n_{03}$).

A special attention is paid to the equation of state. Both the coexistence curve and the spinodal are plotted for the case of a particular parameter $p$ (see appendix~\ref{appC}), which takes a specific value for each system. Analyzing the surface of equation of state (see figure~\ref{fig_11}) we noticed that the curves of pressure sharply increase in the region of high densities (which exceed the liquid phase density). Apparently, the $\rho^4$-model has some boundary density $n_{\text{lim}}$ (see figure~\ref{fig_2}) so that there is a range of physically existent densities $0 \leqslant \bar n \leqslant n_{\text{lim}}$.
 To examine the behavior of the system in the high density region $\bar n \geqslant n_{\text{lim}}$ (see figure~\ref{fig_10}), it might be essential to consider~\eqref{GPF_8} in the approximation of a $\rho^{2m}$-model where $m \geqslant 3$.

 The forthcoming research is to calculate the GPF~\eqref{GPF_9} using the renormalization group method~\cite{ma} with correlation effects taken into account. Thereby, it will be possible to describe the behavior of a simple fluid in a wide temperature range including the critical point within a sole approach.

\section*{Acknowledgements}
This work was partly supported by the European Commission under the project STREVCOMS PIRSES-2013-612669, FP7 EU IRSES projects No.~269139 (DCPPhysBio).

\appendix
\section{Integrating over the coordinates $x_1,\ldots,x_N$ in the GPF} \label{appA}

Now expression of the GPF~\eqref{GPF_2a} becomes slightly different
\begin{align*} \label{GPF_2b}
   \Xi &= {g}_{ \Psi}^{N_v} \int (\rd \rho)^{N_v} \exp \left[ \beta \mu \rho_{0}  +  \frac{\beta}{2} \sum \limits_{\vec{k}\in\cB_c} W(0) \rho_{\vk} \rho_{-\vk} \right] \int (\rd\varphi_{\,\vl})^{N_{v}} \exp  \left( -   \frac{1}{4p} \sum \limits_{\vec{l} \in\Lambda} \varphi_{\,\vl}^2 \right)
    \non
    &\quad\times\int (\rd \nu_{\,\vl})^{N_v} \exp  \left( 2 \piup \ri  \sum \limits_{\vec{l} \in\Lambda} \nu_{\,\vl} \rho_{\,\vl} \right)
     \sum \limits_{N=0}^{\infty} \frac{1}{N!} \int \limits_V (\rd x)^N \exp \left[ - 2 \piup \ri  \sum \limits_{\vec{l} \in\Lambda} \left( \nu_{\,\vl} -  \frac{\varphi_{\,\vl}}{2\piup} \right) \rho_{\,\vl} (\eta) \right],
 \end{align*}
 $\rho_{\,\vl}$ and $\nu_{\,\vl}$ are the representations of the collective variables $\rho_{\vk}$ and $\nu_{\vk}$ in direct space
\[
\rho_{\,\vl} = \frac{1}{\sqrt{N_v}} \sli_{\vec{k}\in\cB_c}\rho_{\vk} \re^{-\ri\vec{k}\vl}, \qquad \nu_{\,\vl} = \frac{1}{\sqrt{N_v}} \sli_{\vec{k}\in\cB_c}\nu_{\vk} \re^{\ri\vec{k}\vl}, \qquad
(\rd\nu_{\,\vl})^{N_v} = \prod \limits_{\vl\in\Lambda} \rd \nu_{\,\vl}\,.
\]
It can be seen from the latter expression that one can integrate over the coordinates of particles in the system. For this purpose, let us consider an integral over $\rd x_1$ in detail
\[
I = \int \limits_V \rd x_1 \exp\left[- 2 \piup \ri   \left( \nu_l -  \frac{\varphi_{\,\vl}}{2\piup} \right) \rho_{\,\vl}  (\eta)\right] = \int \limits_V \rd x_1 \exp\left[- 2 \piup \ri   \left( \nu_{\,\vl} -  \frac{\varphi_{\,\vl}}{2\piup} \right) \sli_{x \in \eta} I_{\Delta_{\,\vl} (x_1)}\right].
\]
 A particle with the coordinate $x_1$ can get into one cell, for example into $\Delta_{\,\vl}$, and cannot get into any other cell, so
\[
I = \int \limits_v \rd x_1 \exp\left[- 2 \piup \ri  \left( \nu_{\,\vl} -  \frac{\varphi_{\,\vl}}{2\piup} \right) \right] = v \exp\left[- 2 \piup \ri  \left( \nu_{\,\vl} -  \frac{\varphi_{\,\vl}}{2\piup} \right) \right].
\]
The rest of integrals over each of $x_i$ give the same contribution. Now, the GPF can be written in the form
\begin{align*}
\Xi &= {g}_{\Psi} \int (\rd \rho_{\,\vl})^{N_v} \exp \left[ \beta \mu \sli_{\vl \in \Lambda}  \rho_{\,\vl} +  \frac{\beta}{2} \sum \limits_{\vec{l}_1, \vec{l}_2 \in\Lambda}  W_{l_{12}} \rho_{\,\vl_1} \rho_{\,\vl_2}  \right]
   \non
   &\quad \times\sum \limits_{m=0}^{\infty} \frac{v^m}{m!} \prod \limits_{\vec{l} \in\Lambda} \int (\rd\varphi_{\,\vl}) \exp\left( -   \frac{1}{4p}  \varphi_{\,\vl}^2 + \ri m \varphi_{\,\vl}\right) \int (\rd \nu_{\,\vl})^{N_v} \exp\left[ 2 \piup \ri  \sum \limits_{\vec{l} \in\Lambda}  (\rho_{\,\vl} - m) \nu_{\,\vl} \right]. \nonumber
 \end{align*}
 Here, $\sqrt{4 \piup p } \, \re^{-pm^2}$ is the result of integrating over $\varphi_{\,\vl}$.

\section{Cumulant representation of the Jacobian of transition } \label{appB}

Let us find an ``entropy'' part of~\eqref{GPF_7} (the Jacobian of transition).  To ease the process, let us use $ J (\tilde t_{\,\vl}) = \prod\limits_{l=1}^{N_v} J_l(\tilde t_{\,\vl})$
\[
J_l(\tilde t_{\,\vl}) = \sli_{m=0}^\infty \frac{v^m}{m!} \re^{-pm^2} \re^{m\tilde t_{\,\vl}}.
\]
This expression can be represented as a cumulant series
\[
\tilde  J_l(\tilde t_{\,\vl}) = \exp \left( - \sli_{n=0}^\infty \frac{a_n}{n!} \tilde t_{\,\vl}^{\,n} \right).
\]
Taking into account the equality $\tilde J_l(\tilde t_{\,\vl})\Big|_{\tilde t_{\,\vl}=0}\equiv J_l(\tilde t_{\,\vl})\Big|_{\tilde t_{\,\vl}=0}$, we have
\[ \label{exp_a0}
\re^{-a_0} = \sli_{m=0}^\infty \frac{v^m}{m!} \re^{-pm^2}.
\]
Applying the following condition
\[ \label{an_condition}
\frac{\partial^n \tilde J_l(\tilde t_{\,\vl})}{\partial \tilde t_{\,\vl}^{\,n}}\Bigg|_{\tilde t_{\,\vl}=0} =
\frac{\partial^n J_l(\tilde t_{\,\vl})}{\partial \tilde t_{\,\vl}^{\,n}}\Bigg|_{\tilde t_{\,\vl}=0},
\]
we can find the cumulants $a_n$ with $n\geqslant 1$. The first four of them are of the form
\begin{align}
\label{an}
a_1= - \frac{T_1 (v, p)}{T_0 (v, p)}\,,& \qquad a_2 = - \frac{T_2 (v, p)}{T_0 (v, p)} + a_1^2\,,
\qquad
a_3 = - \frac{T_3 (v, p)}{T_0 (v, p)} - a_1^3 + 3 a_1 a_2\,,  \non
&a_4= - \frac{T_4 (v, p)}{ T_0 (v, p)} + a_1^4 - 6 a_1^2 a_2 + 4 a_1 a_3 + 3 a_2^2\,,
\end{align}
where the functions $T_n(v, p)$
\[
T_n(v, p) = \sli_{m=0}^\infty \frac{v^m}{m!} m^n \re^{-pm^2}
\]
are rapidly convergent series for all $p>0$. As it can be seen from~\eqref{p1}, the parameter $p$ is proportional to $\Psi(0)$.
The numeric values of these four coefficients in the case of parameters~\eqref{pAzchi_values}
\begin{eqnarray}
a_1 = -0.8668, \qquad
a_2 = -0.4006, \qquad
a_3 = -0.0529, \qquad
a_4 = 0.0299. \nonumber
\end{eqnarray}

\section{Relation between the parameters $p$ and $v$} \label{appC}

Note that the condition $D_{\text b}=0$, [recall that $D_{\text b}$ is expressed by~\eqref{discrim_Db}] yields two real solutions for the density $\bar{n}$ and determines the range of density for the given values of the parameters $p$ and $v$
\[
\bar{n} = n_g \pm \left(  \frac{8 \tilde{a}_2^3}{9 a_4} \right)^{\frac{1}{2}}.
\]
Therefore, we obtain both the minimal and the maximal density
\[
n_{\text{min}} = n_g - \frac{2}{3} \left(  \frac{ 2 \tilde{a}_2^3 }{ a_4 } \right)^{\frac{1}{2}}, \qquad n_{\text{max}} = n_g + \frac{2}{3} \left( \frac{ 2 \tilde{a}_2^3 }{ a_4 } \right)^{\frac{1}{2}}
\]
for the considered cell fluid model.

\begin{figure}[!b]
\centering
\begin{minipage}{0.49\textwidth}
\begin{center}
\includegraphics[width=0.95\textwidth]{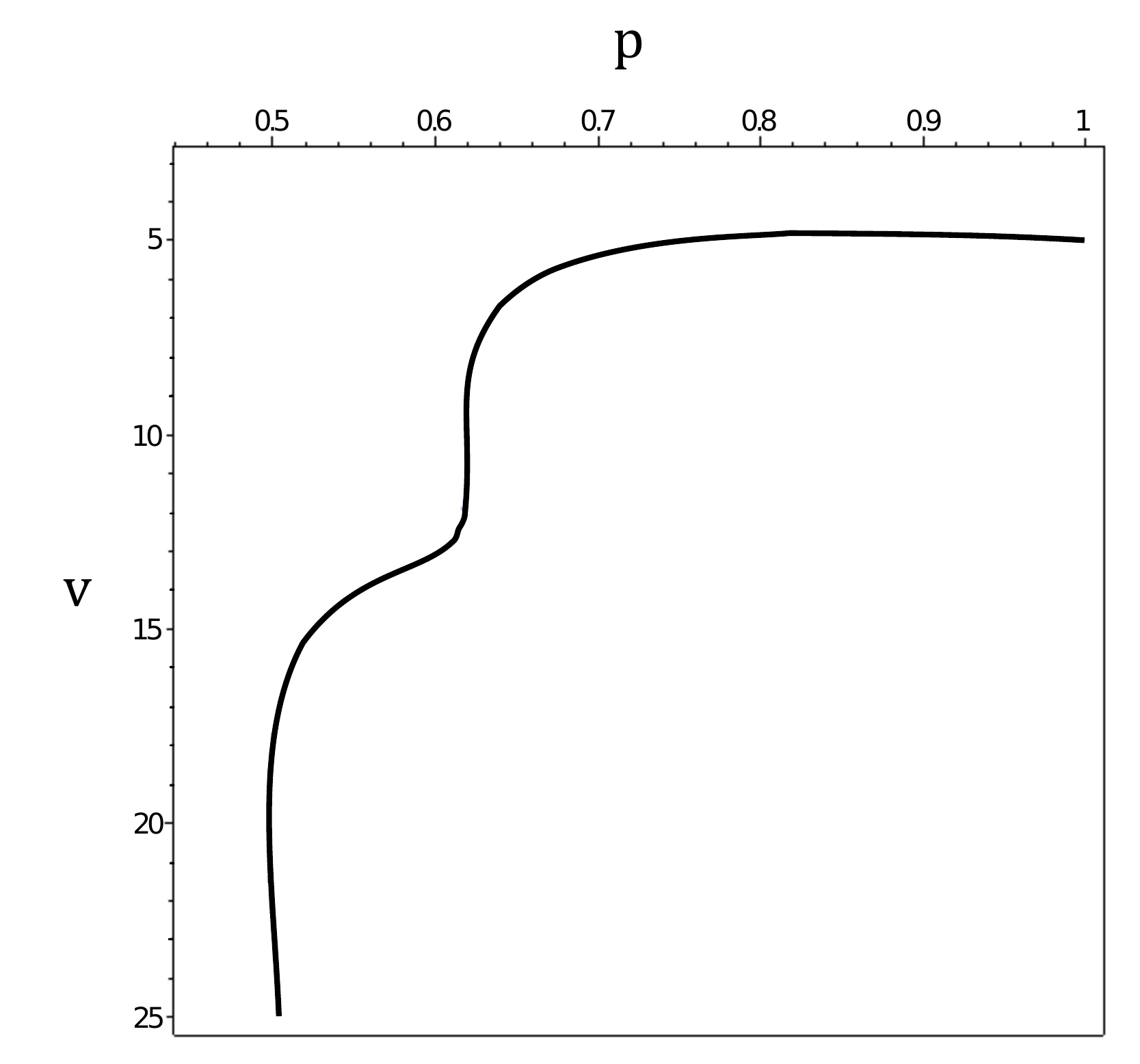}
\caption{Plot of $v$ as a function of $p$.}\label{fig_12}
\end{center}
\end{minipage}
\begin{minipage}{0.49\textwidth}
\begin{center}
\includegraphics[width=0.92\textwidth]{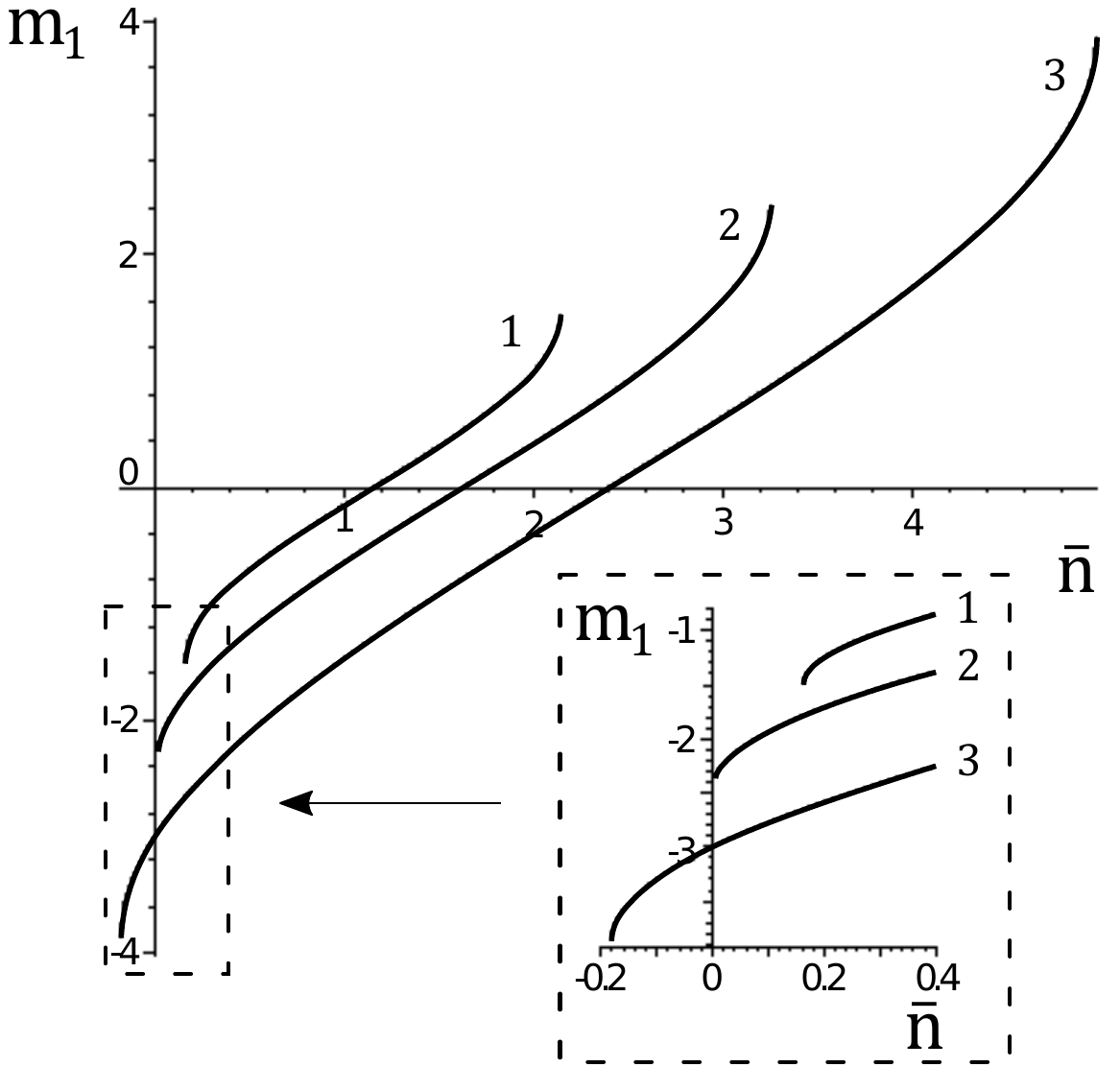}
\caption{Plot of the solution $m_1$~\eqref{solutions_m123} as a function of average density at $p=0.8$ and various values of $v$: $v = 2$ (curve~1), $v =  4.859808$ (curve~2), $v = 6$ (curve~3). }\label{fig_13}
\end{center}
\end{minipage}
\end{figure}

Apparently, $n_{\text{min}}=0$, and hence we have the equation
\[
n_g - \frac{2}{3} \left(  \frac{ 2 \tilde{a}_2^3 }{ a_4 } \right)^{\frac{1}{2}}=0,
\]
allowing us to calculate a respective value of the parameter $v$ (the volume of a cell) for each particular parameter $p$ (see figure~\ref{fig_12}).
Using figure~\ref{fig_13}, it is easy to make sure that not all the values of $v$ in combination with the parameter $p$ would give a correct picture.

Curves $1, 2, 3$ plotted in figure~\ref{fig_13}
match $ p=0.8 $ and three different values of $v$. Curve~2, which is pertinent to $p = 0.8 $ and $v = 4.859808 $, starts at $\bar n=0$. Increasing the value of $v$ at fixed $p$, the least $m_1$ is reached for $\bar{n}<0$ (curve~$3$ for $v= 6 $). Decreasing the value of $v$, the least $m_1$ is for $\bar{n} > 0$ (curve~1 for $v = 2$).

Thus, there is a clear criterion for the selection of values of the parameters $p$ and $v$, which are the arguments of the special function $T_{n}(p,v)$ (see appendix~\ref{appB}). The former $p$ is defined using the formula~\eqref{p1}. This parameter varies upon the type of the investigated substance and depends on the relation $R_0/\alpha$ [for $R_0$ and $\alpha$ see~\eqref{Morse_pot}]. The latter parameter $v$ provides a description of a particular density region of the system  $\bar{n}$ ($ 0 \leqslant \bar{n} \leqslant {n}_{\text{lim}} $, where ${n}_{\text{lim}}$ is the limited density of a liquid phase) and is unambiguously defined by the values of the parameter $p$.

Moreover, $p$ and $v$ should meet the condition $a_4 > 0$. This condition provides a convergency of such integrals as~\eqref{GPF_10}.

This result is very important because all parameters of the model are uniquely linked with each other. So, making estimations for numeric values of the coefficients $a_n$ expressed by (\ref{an}) we set the value of parameter $p$, then find the corresponding value of $v$ using the equation $n_{\text{min}}=0$. Taking into account~\eqref{Tc}, the expression~\eqref{p1} gets the following form
\[
p = \chi \Psi(0) / [ 2 \tilde{a}_2
 W(0,T_{\text c})].
\]
Using the later, one can find the value of $\chi$, which meets the condition $W(k)>0$. Being associated with $p$, the parameter $\chi$ becomes a function of microscopic parameters of an interaction potential, particularly of the relation $R_0/\alpha$.
 In case of the interaction potential $\bar{U}_{l_{12}}=\Psi_{l_{12}}-U_{l_{12}}$ with the Fourier transform~\eqref{fourier_Morse_pot} at $R_0/\alpha=3 \ln 2$, we have the following set of parameters
\begin{equation} \label{pAzchi_values}
p = 0.8, \qquad v = 4.859808, \qquad \chi = 2.5173.
\end{equation}
Recall that length is measured in $R_0$-units.

\ukrainianpart

\title{Фазовий перехід в комірковій моделі плину}
\author{М.П. Козловський, О.А. Добуш}
\address{Інститут фізики конденсованих систем НАН України, вул. Свєнціцького, 1, 79011 Львів, Україна}

\makeukrtitle

\begin{abstract}
\tolerance=3000%
Запропоновано метод опису фазового переходу в комірковій моделі плину із парним потенціалом взаємодії, що містить частини відштовхування і притягання.
Отримано точне представлення великої статистичної суми цієї моделі в формалізмі колективних змінних.
У наближенні типу молекулярного поля досліджено поведінку системи при температурах вищих і нижчих, ніж критична.
Використовуючи рівняння, що зв’язує хімічний потенціал  та густину, розраховано явну аналітичну форму рівняння стану, що справедливе для широкої області температур.
Представлено криву співіснування, поверхню рівняння стану та діаграму стану коміркового плину Морзе.

\keywords колективні змінні, коміркова модель, плин Морзе, рівняння стану, фазовий перехід

\end{abstract}


\begin{thebibliography}{99}
\bibitem{hansen}
   Hansen J.-P., McDonald I.R., Theory of Simple Liquids: with Applications to Soft Matter, 4th Edn., Academic Press, Oxford, 2013.

\bibitem{PhysRevE.85.031131}
Bertrand C.E., Nicoll J.F., Anisimov M.A., Phys. Rev. E, 2012, \textbf{85}, 031131, \bibdoi{10.1103/PhysRevE.85.031131}.

\bibitem{Anisimov_CMP_13}
Anisimov M.A., Condens. Matter Phys., 2013, \textbf{16}, 23603, \bibdoi{10.5488/CMP.16.23603}.

\bibitem{Pini_Stell_Wilding_98}
Pini D., Stell G., Wilding N.B., Mol. Phys., 1998, \textbf{95}, 483, \bibdoi{10.1080/00268979809483183}.
	
\bibitem{Lee_Stell_Hoye_04}
Lee C.-L., Stell G., H\o ye J., J. Mol. Liq., 2004, \textbf{112}, 13, \bibdoi{10.1016/j.molliq.2003.11.004}.
	
\bibitem{Par_Rea_12}
Parola A., Reatto L., Mol. Phys., 2012, \textbf{110}, 2859, \bibdoi{10.1080/00268976.2012.666573}.

\bibitem{Caillol_06}
Caillol J.-M., Mol. Phys., 2006, \textbf{104}, 1931, \bibdoi{10.1080/00268970600740774}.
	
\bibitem{virial}
Schultz A.J., Kofke D.A., Fluid Phase Equilib., 2016, \textbf{409}, 12, \bibdoi{10.1016/j.fluid.2015.09.016}.

\bibitem{yukhn_2014}
Yukhnovskii I.R., Condens. Matter Phys., 2014, \textbf{17}, 43001, \bibdoi{10.5488/CMP.17.43001}.

\bibitem{tang}
Tang Y., J. Chem. Phys., 2011, \textbf{134}, 224508, \bibdoi{10.1063/1.3599048}.
	
\bibitem{rebenko_2013}
Rebenko A.L., Rev. Math. Phys., 2013, \textbf{25}, 1330006,  \bibdoi{10.1142/S0129055X13300069}.

\bibitem{rebenko_2011}
Petrenko S.M., Rebenko A.L., Tretychnyi M.V., Ukr. Math. J., 2011, \textbf{63}, 425, \bibdoi{10.1007/s11253-011-0513-0}.

\bibitem{apf_11}
Apfelbaum E.M., J. Chem. Phys., 2011, \textbf{134}, 194506, \bibdoi{10.1063/1.3590201}.

\bibitem{martinez}
Mart\'{i}nez-Valencia A., Gonz\'{a}lez-Melchor M., Orea P., L\'{o}pez-Lemus J., Mol. Simul., 2013, \textbf{39}, 64,\\ \bibdoi{10.1080/08927022.2012.702422}.

\bibitem{okumura_00}
Okumura H., Yonezawa F., J. Chem. Phys., 2000, \textbf{113}, 9162, \bibdoi{10.1063/1.1320828}.

\bibitem{singh}	
Singh J.K., Adhikari J., Kwak S.K., Fluid Phase Equilib., 2006, \textbf{248}, 1, \bibdoi{10.1016/j.fluid.2006.07.010}.

\bibitem{KK_arxiv}
Kozitsky Yu., Kozlovskii M., Preprint \arxiv{1610.01845}, 2016.

\bibitem{yukhn_1980}
Yukhnovskii I.R., Holovko M.F., Statistical Theory of Classical Equilibrium Systems, Naukova Dumka, Kiev, 1980, (in Russian).

\bibitem{KD_2016}
Kozlovskii M., Dobush O., J. Mol. Liq., 2016, \textbf{215}, 58, \bibdoi{10.1016/j.molliq.2015.12.018}.

\bibitem{girifalko}
Girifalko I.A., Weizer V.G., Phys. Rev., 1959, \textbf{114}, 687, \bibdoi{10.1103/PhysRev.114.687}.

\bibitem{lincoln}
Lincoln R.C., Koliwad K.M., Ghate P.B., Phys. Rev., 1967, \textbf{162}, 854, \bibdoi{10.1103/PhysRev.162.854.2}.

\bibitem{bringas}
Bringas J.G., L\'{o}pez-Lemus J., Ibarra-Tandi B., Orea P., Mol. Simul., 2011, \textbf{37}, 449,\\ \bibdoi{10.1080/08927022.2011.551883}.

\bibitem{KR_2010}
Kozlovskii M., Romanik R., Condens. Matter Phys., 2010, \textbf{13}, 43004, \bibdoi{10.5488/CMP.13.43004}.

\bibitem{KR_2011}
Kozlovskii M., Romanik R., Condens. Matter Phys., 2011, \textbf{14}, 43002, \bibdoi{10.5488/CMP.14.43002}.

\bibitem{KR_2012}
Kozlovskii M., Romanik R., J. Mol. Liq., 2012, \textbf{167}, 14, \bibdoi{10.1016/j.molliq.2011.12.003}.

\bibitem{ma}
Ma Sh.-K., Rev. Mod. Phys., 1973, \textbf{45}, 589, \bibdoi{10.1103/RevModPhys.45.589}.

\bibitem{K_2009}
Kozlovskii M.P., Condens. Matter Phys., 2009, \textbf{12}, 151, \bibdoi{10.5488/CMP.12.2.151}.

\bibitem{KPP_p_2006}
Kozlovskii M.P., Pylyuk I.V., Prytula O.O., Phys. Rev. B, 2006, \textbf{73}, 174406, \bibdoi{10.1103/PhysRevB.73.174406}.

\bibitem{K_2007}
Kozlovskii M.P., Phase Transitions, 2007, \textbf{80}, 3, \bibdoi{10.1080/01411590701315161}.

\bibitem{yukhn_2013}
Yukhnovskii I., Kolomiets V., Idzyk I., Condens. Matter Phys., 2013, \textbf{16}, 23604, \bibdoi{10.5488/CMP.16.23604}.

\bibitem{kadanoff}
Kadanoff L.P., J. Stat. Phys., 2009, \textbf{137}, 777, \bibdoi{10.1007/s10955-009-9814-1}.

\bibitem{bulavin}
Bulavin L.A., Kulinskii V.L., J. Chem. Phys., 2010, \textbf{133}, 134101, \bibdoi{10.1063/1.3496468}.


\end{thebibliography}
\end{document}